\documentclass[aps,amssymb,11pt,showpacs,showkeys,letterpaper]{revtex4-1}
\usepackage{amsmath, amssymb}
\usepackage{graphicx}
\usepackage{epsfig}
\usepackage{bm}

\newcommand{\be}{\begin{equation}}
\newcommand{\ee}{\end{equation}}
\newcommand{\beq}{\begin{eqnarray}}
\newcommand{\eeq}{\end{eqnarray}}

\begin{document}

\title{Testing the EoS of dark matter with cosmological observations}

\author{Arturo Avelino}
\email[]{avelino@fisica.ugto.mx}
\affiliation{Departamento de F\'isica, DCI, Campus Le\'on, Universidad de
Guanajuato,  CP. 37150, Le\'on, Guanajuato, M\'exico.}

\author{Norman Cruz}
\email[]{norman.cruz@usach.cl}
\affiliation{\it Departamento de F\'{\i}sica, Universidad
de Santiago de Chile, Casilla 307,
Santiago , Chile.}

\author{Ulises Nucamendi}
\email[]{ulises@ifm.umich.mx}
\affiliation{Instituto de F\'{\i}sica y Matem\'aticas,
Universidad Michoacana de San Nicol\'as de Hidalgo \\
Edificio C-3, Ciudad Universitaria, CP. 58040,
Morelia, Michoac\'an, M\'exico.}


\begin{abstract}
We explore the cosmological constraints on the parameter
$w_{\rm dm}$ of the dark matter barotropic equation of state (EoS) to
investigate the ``warmness'' of the dark matter fluid.
The model is composed by the dark matter and dark energy fluids in addition to
the radiation and baryon components.
We constrain the values of $w_{\rm dm}$ using the latest cosmological
observations that measure the expansion history of the Universe. When $w_{\rm
dm}$ is estimated together with the parameter $w_{\rm de}$ of the
barotropic EoS of dark energy we found that the cosmological data favor a
value of $w_{\rm dm} = 0.006 \pm 0.001$,  suggesting a \textit{warm} dark
matter, and $w_{\rm de}= -1.11 \pm 0.03$ that corresponds to a phantom dark
energy, instead of favoring a cold dark matter and a cosmological constant
($w_{\rm dm}=0, w_{\rm de} = -1$).
When $w_{\rm dm}$ is estimated alone but assuming $w_{\rm de} =
 -1, -1.1, -0.9$, we found $w_{\rm dm} = 0.009 \pm 0.002$, $0.006
\pm 0.002 $, $0.012 \pm 0.002$ respectively, where the errors are at
3$\sigma$ (99.73\%), i.e., $w_{\rm dm} > 0$ with at least 99.73\% of
confidence level. 
When $(w_{\rm dm}, \Omega_{\rm dm0})$ are constrained together, the best fit
to data corresponds to $(w_{\rm dm}=0.005 \pm 0.001$, $\Omega_{\rm dm0} = 
0.223 \pm 0.008)$ and with the assumption of $w_{\rm de} = -1.1$ instead of a
cosmological constant (i.e., $w_{\rm de} = -1$).
With these results we found evidence of $w_{\rm dm} > 0$ suggesting a
\textit{warm} dark matter, independent of the assumed value for $w_{\rm de}$,
but where values $w_{\rm de} < -1$ are preferred by the observations instead
of
the cosmological constant. These constraints on $w_{\rm dm}$ are consistent
with perturbative analyses done in previous works.
\end{abstract}


\maketitle

\section{\label{sec:Int}Introduction}

The astrophysical  evidence for the existence of Dark Matter (DM)
is well based on observations from the scales of galaxies,
clusters and the universe itself, in the framework of the standard
cosmological model.

Despite the fact that the cosmological scenario where cosmological
parameters fit a dark matter mainly non-baryonic and cold, a
great debate has currently opened about the possibility that  Warm
Dark Matter becomes a better candidate to understand the recent
investigations. For a wide discussion based in the new results in
the area see~\cite{workshop}. A summary of astrophysical
constraints on dark matter is present in~\cite{Tao}

Let us summarize some of the difficulties of the Cold Dark Matter (CDM)
model. At galactic scales, N-body simulations of cosmological
structures with CDM have predicted that the dark matter halos
surrounding galaxies must present radial profiles of the mass
density and velocity dispersion with a central cusp in which
the value of the logarithmic slope is under discussion (see
~\cite{Hernquist}).


In the case of the {\it missing satellites problem
}~\cite{Klypin},~\cite{Moore} exists a discrepancy in the cold dark
matter model between the predicted numbers of satellite galaxies
inside the galactic halo for the Milky Way and lower number
observed.  This problem has been undertaken assuming a warm dark
matter component in various works~\cite{WDMworks}.

A recent investigation of N-body simulations in a warm dark matter
scenario, which took into account the new dwarf spheroidal
galaxies discovered in the Sloan Digital Sky Survey
(SDSS)~\cite{Castander}, derived lower limits on the dark matter
particle mass~\cite{Polisensky}.



In recent investigations, the measuring of the dark matter equation
of state (EoS) has been carried out using different approaches.
Following a suggestion given in \cite{Faber}, where the method
combines kinematic and gravitational lensing data, the dark
matter EoS was measured in \cite{Serra} using
galaxy clusters which present gravitational lensing effects. The
result of this work indicates that the measured EoS for dark
matter is consistent with the standard pressureless cold dark
matter at 1$\sigma$ level. Nevertheless, lensing analysis in
clusters such as Coma and CL0024 shows a trend to prefer an exotic
EoS for the dark matter, i.e., $w \sim - 1/3$.



Models of the dark matter component described by a fluid with
non-zero effective pressure has been studied in some astrophysical
scenarios.  At galactic level, an EoS with anisotropic pressures has
been investigated in \cite{Bhara} in order to explain flat
rotation curves. A polytropic dark matter halo fits very well a number
of elliptical galaxies, improving or at least giving similar
results to the velocity dispersion profile compared to a
stars-only model \cite{Bhara}.


Explorations of the EoS for dark matter at cosmological level have been carried out
in various frameworks. In \cite{Muller}, a constant EoS for dark matter is
studied from the study of the power spectrum, assuming a cosmological constant as the
dark energy fluid and a flat universe. The bounds obtained for $w_{\rm dm}$
were $-1.50\times 10^{-6}<w_{\rm dm}<1.13\times 10^{-6}$ if there is no
entropy production and $-8.78\times 10^{-3}<w_{\rm dm}<1.86\times 10^{-3}$
if the adiabatic sound speed vanishes. Phenomenologically,
EoS for both dark fluids have been studied in \cite{Kumar} using 
WMAP+BAO+$H_{O}$
observations by
synchronizing the model with the $\Lambda CDM$ model at the present time. The
dark matter component behaves like radiation at very early times and
at the present time $w_{\rm dm}=0.0005$.

In the case of unified dark matter models, where a single matter component
is assumed to source the acceleration and structure formation
\cite{Bertacca}, the initial phase is described by a cold dark matter so that
the fitting with cosmological data leads
to a late phase with negative $w_{\rm dm}$ very close to a cosmological
constant or phantom matter.

Our aim in this work is to study the EoS of the dark matter
component allowing a non zero value for $w_{\rm dm}$ from the beginning
and then to undertake a constraining of its value using the latest observations
that measure the expansion history of the universe. In what follows,
we will assume a barotropic EoS for this component.
Of course, this assumption is rather restrictive because in approaches
based in the nature of particles constituting the dark matter
fluid is expected to have a $w_{\rm dm}$ varying with the cosmological time.
Such is the case, for example, for dark matter Bose-Einstein condensation
\cite{Harko}.
Nevertheless, if the dark matter fluid is modeled, in the non-relativistic approximation,
as a non-degenerated ideal Maxwell-Boltzmann gas, a barotropic EoS is obtained
with $w_{\rm dm}=$ constant \cite{HarkoandLobo}.

The present paper is organized as follows. In Section II, we briefly outline
the basic equations of evolution of the model. In Section III, the parameters
of the model are constrained using cosmological data from type Ia supernovae,
CMBR, baryon acoustic oscillations, the Hubble expansion rate and the age of the
universe.
Finally, in Section IV, we discuss and conclude our results.

\section{\label{sec:kke}The cosmological model}

We study a cosmological model composed by four fluids: radiation, baryons,
dark matter and dark energy. We assume a barotropic equation of state
(EoS) for dark matter (dm) and energy (de) fluids, $p_{i}= w_{i} \cdot
\rho_{i}$, with $i = $ dm, de, respectively. $\rho_i$ corresponds to
the density of the fluid and $p_i$ to its pressure.
We are interested in studying the cosmological prediction for the
EoS of the dark matter, in particular, for the magnitude of $w_{\rm dm}$.

We assume a spatially flat Friedmann-Robertson-Walker (FRW) cosmology.
The Friedmann constraint and conservation equations for the radiation,
baryonic, dark matter and dark energy fluids are given respectively as

\begin{align}
H^2 = & \frac{8 \pi G}{3} \left(\rho_{\rm r} + \rho_{\rm b} +
\rho_{\rm de} + \rho_{\rm dm} \right) \label{FriedmannEq1st} \\
0 = & \dot{\rho}_{\rm r} + 4H \rho_{\rm r} \label{ConserEqRadiation} \\
0 = & \dot{\rho}_{\rm b} + 3H \rho_{\rm  b} \label{ConserEqBaryon} \\
0 = & \dot{\rho}_{\rm dm} + 3H \rho_{\rm dm} (1+w_{\rm dm})
\label{ConserEqDarkMatter} \\
0 = & \dot{\rho}_{\rm de} + 3H \rho_{\rm de} (1+w_{\rm de}),
\label{ConserEqDarkEnergy}
\end{align}

\noindent where $H$ is the Hubble parameter and the dot over
$\dot{\rho}_i$ stands for the derivative with respect to the cosmic time.
The conservation equations
(\ref{ConserEqRadiation})-(\ref{ConserEqDarkMatter}) have the respective
solutions in terms of the scale factor $a$

\begin{equation}\label{DensitiesExpressions}
\rho_{\rm r}(a) = \frac{\rho_{\rm r0}}{a^4}, \; \; \; \; \; \; \rho_{\rm
b}(a) = \frac{\rho_{\rm r0}}{a^3}, \; \; \; \; \; \; \rho_{\rm dm}(a) =
\frac{\rho_{\rm dm0}}{a^{3(1+w_{\rm dm})}},
\; \; \; \; \; \; \rho_{\rm de}(a) = \frac{\rho_{\rm de0}}{a^{3(1+w_{\rm
de})}},
\end{equation}

\noindent where the subscript zero at $\rho_{i0}$ indicates the
present-day values of the respective matter-energy densities.
Inserting the expression
(\ref{DensitiesExpressions}) on the Friedmann constraint
(\ref{FriedmannEq1st}), and dividing by the Hubble constant $H_0$, it becomes

\begin{equation}\label{HubbleParameterDensitiesScaleFactor}
E^2(a) \equiv \frac{H^2(a)}{H^2_0} = \frac{8 \pi G}{3 H^2_0} \left(
\frac{\rho_{\rm r0}}{a^4} + \frac{\rho_{\rm b0}}{a^3} +\frac{\rho_{\rm
dm0}}{a^{3(1+w_{\rm dm})}}  + \frac{\rho_{\rm de0}}{a^{3(1+w_{\rm de})}}
\right).
\end{equation}

We define the dimensionless parameter densities as $\Omega_{i0} \equiv
\rho_{i0}/\rho^0_{\rm crit}$, where $\rho^0_{\rm crit}$ is the critical
density evaluated today defined as $\rho^0_{\rm crit} \equiv 3H^2_0
/(8 \pi G)$. With this definition, the Friedmann equation
(\ref{HubbleParameterDensitiesScaleFactor}) obtains the form

\begin{equation}\label{HubbleParameterScaleFactor}
E^2(a)  = \frac{\Omega_{\rm r0}}{a^4} + \frac{\Omega_{\rm
b0}}{a^3} + \frac{\Omega_{\rm dm0}}{a^{3(1+w_{\rm
dm})}} + \frac{\Omega_{\rm de0}}{a^{3(1+w_{\rm de})}},
\end{equation}

\noindent or, using the relation between the scale factor and the redshift
``$z$'' given by  $a = 1/(1+z)$, we rewrite the dimensionless eq.
(\ref{HubbleParameterScaleFactor}) in terms of the redshift as

\begin{equation}\label{HubbleParameterRedshift}
E^2(z)  = \Omega_{\rm r0}(1+z)^4 + \Omega_{\rm b0}(1+z)^3 + \Omega_{\rm
dm0}(1+z)^{3(1+w_{\rm dm})} + \Omega_{\rm de0}(1+z)^{3(1+w_{\rm de})}.
\end{equation}

Setting $E(z=0) =1$ we have the constraint equation,
\begin{equation}
\Omega_{\rm de0} = 1- (\Omega_{\rm r0} + \Omega_{\rm b0} +
\Omega_{\rm dm0}).
\end{equation}

\section{Cosmological constraints}


To constrain the value of $ w_{\rm dm}$ using
cosmological data, to compute their confidence intervals and to calculate their best
estimated values, we use the following cosmological observations described
below measuring the expansion history of the Universe.

To perform the numerical calculations, it was used for the baryonic and
radiation (photons and relativistic neutrinos) components the values of
$\Omega_{b0} = 0.0458$ \cite{WMAP7yKomatsu2011} and $\Omega_{r0} = 0.0000766$
respectively, where the later value is computed from the expression
\cite{WMAP5yKomatsu2009}

\begin{equation}
\Omega_{\rm r0} = \Omega_{\gamma 0} (1+0.2271 N_{\rm eff})
\end{equation}

\noindent where $N_{\rm eff} = 3.04$ is the number of standard neutrino species
\cite{WMAP7yKomatsu2011,SDSS7yReid2010} and $\Omega_{\gamma 0} = 2.469
\times 10^{-5} h^{-2}$ corresponds to the present-day photon density
parameter for a temperature of $T_{\rm cmb} = 2.725$ K
\cite{WMAP7yKomatsu2011}, where $h$ is the dimensionless Hubble
constant  $h \equiv H_0 /(100$ km/s$\cdot$Mpc).

	\subsubsection{Type Ia Supernovae}

We use the type Ia supernovae (SNe Ia) of the ``Union2.1'' data set
(2012) from the Supernova Cosmology Project (SCP) composed of 580 SNe
Ia \cite{Union2.1:Suzuki2011}.
The luminosity distance $d_L$ in a spatially flat FRW Universe is defined
as

\begin{equation}
d_L(z, w_{\rm dm}) = \frac{c(1+z)}{H_0}
\int_0^z
\frac{dz'}{E(z', w_{\rm dm})}
\end{equation}

\noindent where ``$c$'' corresponds to the speed of light in units of
km/sec. The theoretical distance moduli $\mu^t$ for the k-th supernova
at a distance $z_k$ is given by

\begin{equation}
\mu^t(z, w_{\rm dm}) = 5 \log \left[
\frac{d_L(z, w_{\rm dm}) }{\rm Mpc} \right] + 25
\end{equation}

So, the $\chi^2$ function for the SNe Ia test is defined as

\begin{equation}\label{Chi2FunctionSNe}
\chi^2_{\rm SNe}(w_{\rm dm}, H_0) \equiv \sum_{k=1}^n
\left( \frac{\mu^t(z_k, w_{\rm dm}, H_0)-
\mu_k}{\sigmạ_k} \right)^2
\end{equation}

\noindent where $\mu_k$ is the observed distance moduli of the k-th
supernova, with a standard deviation of $\sigmạ_k$ in its
measurement, and $n= 580$.
It was used a \textit{constant} prior distribution function for $H_0$ to
marginalize it (i.e., it is not assumed any particular value of $H_0$) 
because $H_0$ is a nuisance parameter in the SNe Ia test.


\begin{table}
  \centering
\begin{tabular}{ c  | c c | c c }

\multicolumn{5}{c}{\textbf{Best estimates for $(w_{\rm dm}, w_{\rm de})$} }\\


Data set &  $w_{\rm dm}$ &  $w_{\rm de}$ & $\chi^2_{{\rm min}}$ &
$\chi^2_{{\rm d.o.f.}}$ \\
\hline
\hline

SNe Ia & $0.006^{+0.133}_{-0.096}$ & $-1.003^{+0.12}_{-0.13}$  & $ 562.23$ &
$0.97$ \\

$(\mathcal{R}, l_A, z_*)$ CMB & $ 0.004 \pm 0.001$  &
$-1.197^{+0.057}_{-0.053}$ & 1.11 & 1.11 \\

$H(z)$  & $0.007^{+0.069}_{-0.058}$ & $ -1.197^{+0.14}_{-0.13}$ & $8.05$ &
$0.73$ \\

\hline
SNe + CMB + BAO + $H(z)$  & $0.006 \pm 0.001$ & $-1.115 \pm 0.033$ & 578.84 &
0.97 \\
\hline

\end{tabular}
\caption{Best estimated values for $(w_{\rm dm}, w_{\rm de})$. See figures
\ref{PlotGroupWdmWde} and \ref{PlotGroupWdmWdeZoom} for
the confidence intervals. 
We find that
the cosmological data used in the present work favor a non-vanishing
magnitude, positive value for $w_{\rm dm}$ suggesting a \textit{warm} dark
matter, in addition to $w_{\rm de} < -1$ indicating a \textit{phantom} dark
energy. 
In order to compare these results with the $\Lambda$CDM model we computed the
value of the $\chi^2$ function evaluated at $(w_{\rm dm} =0, w_{\rm de}=-1)$
using the same four cosmological data sets (SNe + CMB + BAO + $H(z)$)
together, finding a value of $\chi^2_{\rm \Lambda CDM} = 740.5$, that is
clearly greater
than $\chi^2_{\rm min} = 578.8$ obtained in the present work for $w_{\rm
dm}=0.006$, $w_{\rm de} = -1.115$, indicating that the
$\Lambda$CDM model fits not too well the cosmological data compared with
the latter values.
It was assumed $\Omega_{\rm b0} = 0.0458$, $\Omega_{\rm r0} = 0.0000758$, 
$\Omega_{\rm dm0} = 0.23$ and $H_0 = 73.8$ km/(s$\cdot$Mpc).
The errors correspond to 68.3\% of confidence level ($1\sigma$).}
\label{TableWdmWde}
\end{table}



\begin{table}
  \centering
\begin{tabular}{ c |  c  c c }

\multicolumn{4}{c}{\textbf{Best estimates for $w_{\rm dm}$}}\\


 $w_{\rm dm}$ &  $w_{\rm de}$ & $\chi^2_{{\rm min}}$ &
$\chi^2_{{\rm d.o.f.}}$ \\
\hline \hline


$0.009 \pm 0.002 $ & $-1$ & 591.57 & 0.99 \\
$0.006 \pm 0.002$ & $-1.1$ & 579.02 &  0.97 \\
$0.012 \pm 0.002$ & $-0.9$ & 628.21 & 1.05 \\

\hline

\end{tabular}
\caption{Best estimated values for $w_{\rm dm}$ when it is assumed the
different values of $w_{\rm de} = -1, -1.1, -0.9$ for the dark energy. In the
three cases it is found a non-vanishing positive value for $w_{\rm dm}$.
We find also that the best fit to data is for the case when it is assumed
$w_{\rm de} = -1.1$, i.e., it has the smallest value of $\chi^2_{\rm min}$
compared with the other two cases.
It was used the joint SNe + CMB + BAO + $H(z)$ data sets.
The errors are at 99.73\% of confidence level ($3\sigma$). See figure
\ref{PlotGaussWdmAllTogetherJoinU21FDM} for the likelihood functions. } 
\label{TableWdmAlone}
\end{table}



\begin{table}
  \centering
\begin{tabular}{ c  | c c c | c c }
\multicolumn{6}{c}{\textbf{Best estimates for $(w_{\rm dm}, \Omega_{\rm
dm0})$}}\\


Data set &  $w_{\rm dm}$ &  $\Omega_{\rm dm0}$ &  $w_{\rm de}$ & 
$\chi^2_{{\rm min}}$ & $\chi^2_{{\rm d.o.f.}}$ \\
\hline
\hline

 & $0.004 \pm 0.27$   & $0.229^{+0.16}_{-0.09}$ & $-1$ & $562.22$   & 
0.972 \\
SNe Ia & -0.103 & 0.316 & $-1.1$ & 562.20 & 0.972 \\
& 0.177 & 0.138 & $-0.9$ & 562.275 & 0.972 \\
\hline

 & $-0.0009 \pm 0.003$  & $0.183^{+0.010}_{-0.009}$ & $-1$ & 2.558 & 2.558 \\
$(\mathcal{R}, l_A, z_*)$ CMB & 0.001  & 0.206 & $-1.1$ & 0.04 & 0.04 \\
 & -0.007 & 0.153 & $-0.9$ & 10.90 & 10.90 \\
\hline

& $0.215^{+0.226}_{-0.212}$  &  $0.114^{+0.076}_{-0.050}$  & $-1$ &
$8.271$ & 0.751 \\
$H(z)$   & 0.085  & 0.176  & $-1.1$ & 8.12  & 0.738 \\
   & 0.426 & 0.057 & $-0.9$ & 8.55 & 0.777 \\
\hline \hline

 & $0.005 \pm 0.002$   &  $0.204 \pm 0.008$ & $-1$ & $582.11$ & 0.978 \\
SNe + CMB + BAO + $H(z)$  & $0.005 \pm 0.001$ & $0.223 \pm 0.008$ & $-1.1$ &
578.27 & 0.971 \\
 & $0.005 \pm 0.002$ & $0.184 \pm 0.007$ & $-0.9$ & 601.41 & 1.010 \\
\hline \hline
\end{tabular}
\caption{Best estimated values of the parameter density of dark matter
$\Omega_{\rm dm0}$ and the $w_{\rm dm}$ of the EoS of dark matter $(p_{\rm
dm} = w_{\rm dm} \cdot \rho_{\rm dm})$.  
The first column shows the cosmological data sets used to compute the best
estimates shown in second and third columns. The fourth column 
indicates the assumed value for $w_{\rm de}$. The fifth and sixth
columns correspond to the minimum value of the $\chi^2$ function, $\chi^2_{\rm
min}$, and $\chi^2$ by degrees of freedom, $\chi^2_{\rm d.o.f.}$ respectively.
The latter is defined as $\chi^2_{\rm d.o.f.}= \chi^2_{\rm min} / (n-p)$,
where $n$ is the number of data and $p$ the number of free parameters (in
this case $p=2$).
The computed values come from the minimization of the $\chi^2$
functions defined in (\ref{Chi2FunctionSNe}), (\ref{Chi2FunctionCMB3p}),
(\ref{Chi2FunctionHz}) and (\ref{Chi2FunctionTotal}) respectively.
The fourth row (SNe + CMB + BAO + $H(z)$) encloses the information of all the
cosmological
observations used in the present work to constrain the values of
$(\Omega_{\rm dm0},w_{\rm dm})$. Notice that $w_{\rm dm}$ has a positive
value, favoring a \textit{warm} instead of a cold dark matter.
See figures \ref{PlotsCIs} to \ref{PlotsOdmWdmWdeM09Zoom} for
the confidence intervals.}               
\label{TableWdmOdm}
\end{table}



\begin{figure}
\begin{center}
\hfill%
\includegraphics[width=7cm]{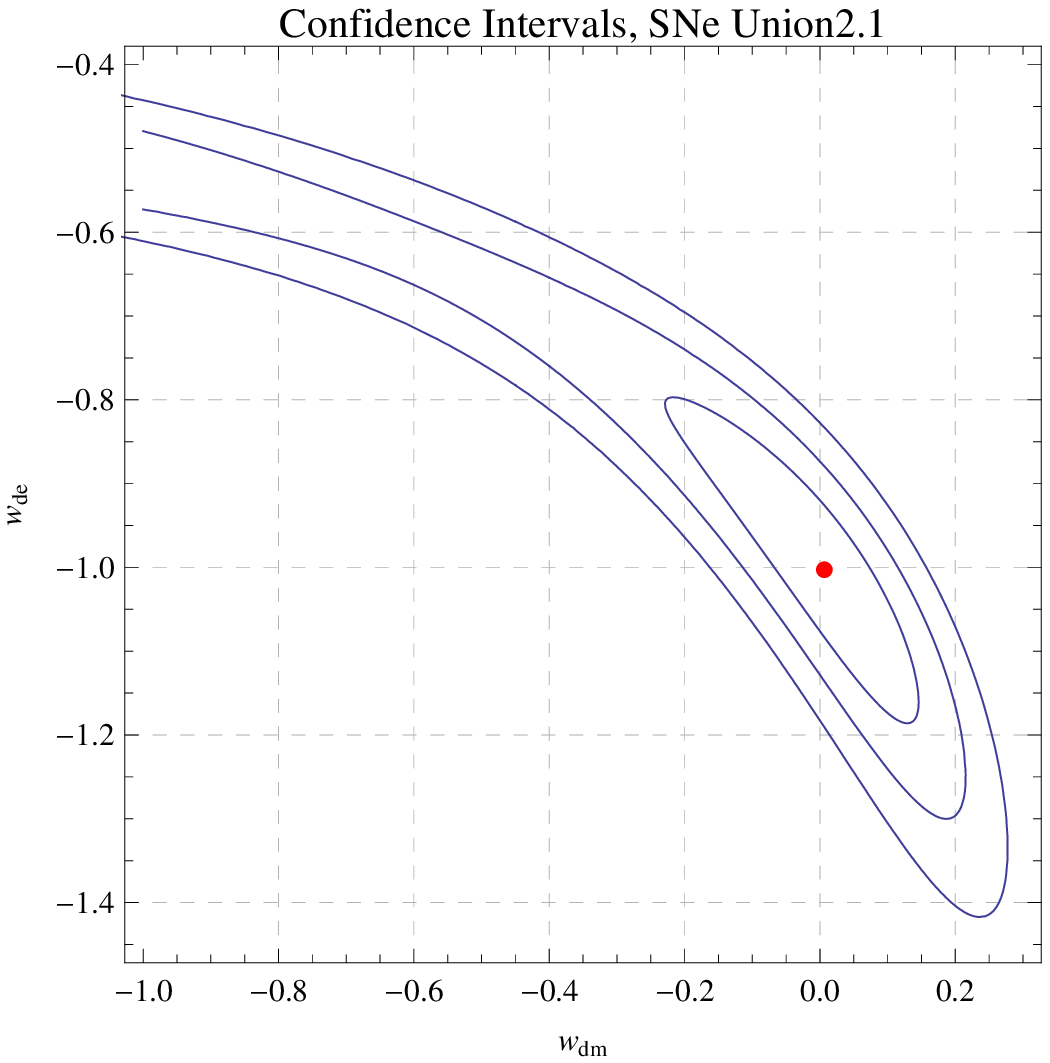}%
\hfill%
\includegraphics[width=7cm]{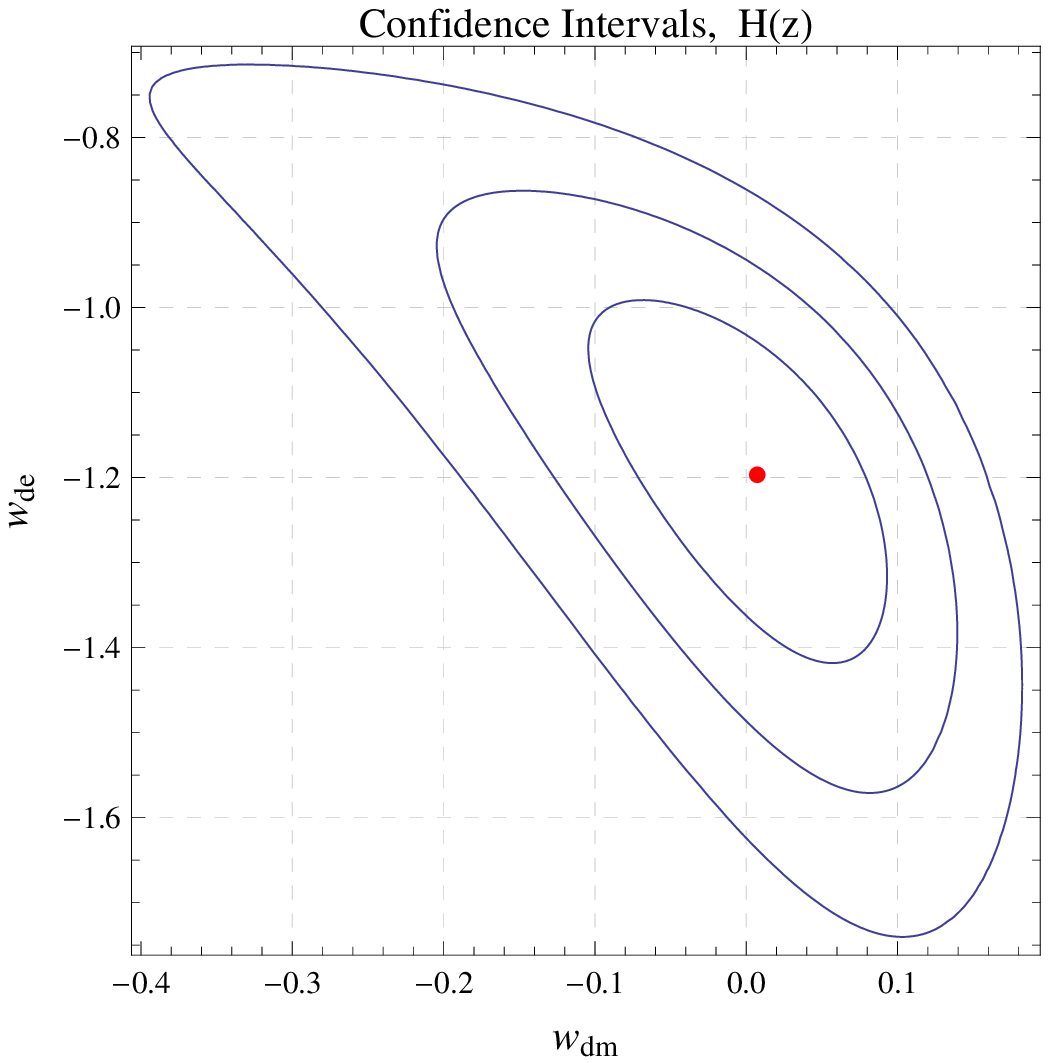}%
\hspace*{\fill}

\hfill%
\includegraphics[width=7cm]{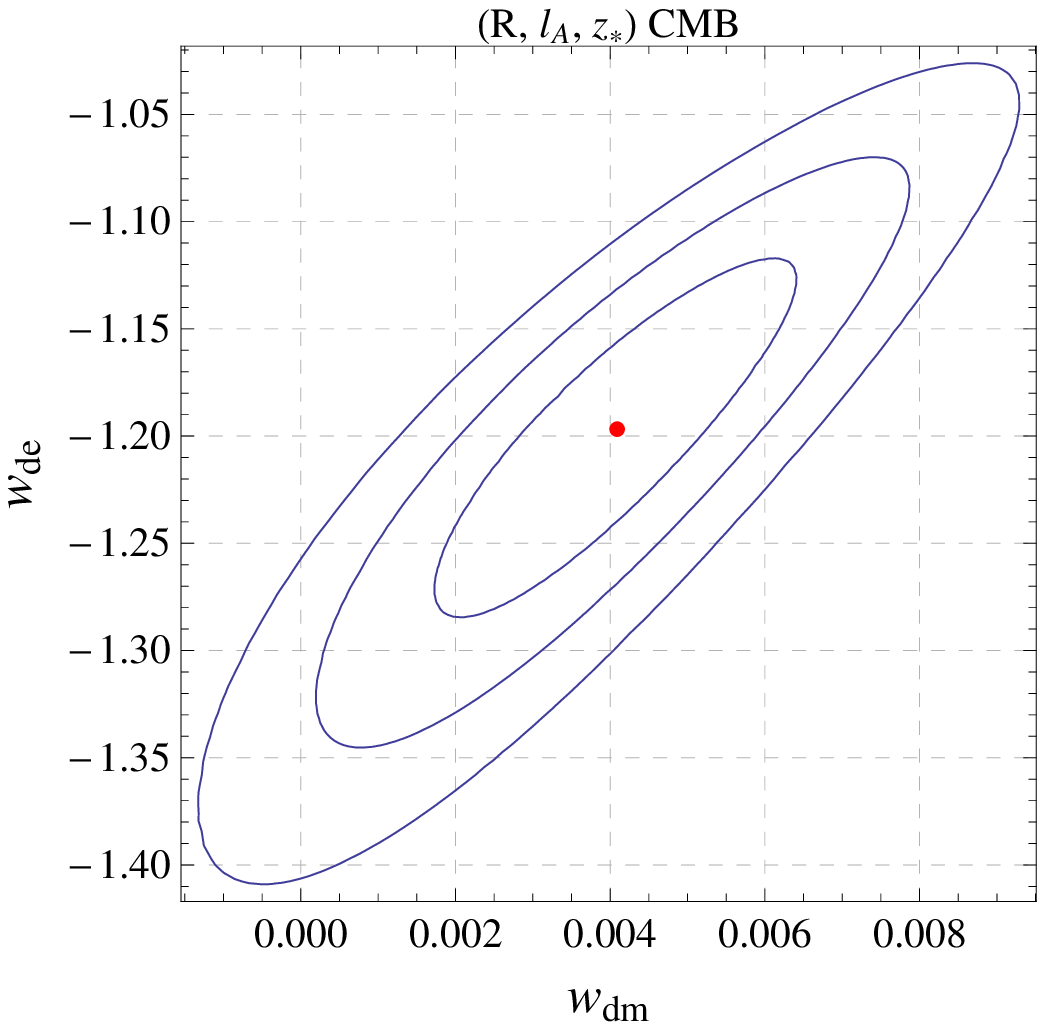}%
\hfill%
\includegraphics[width=7cm]{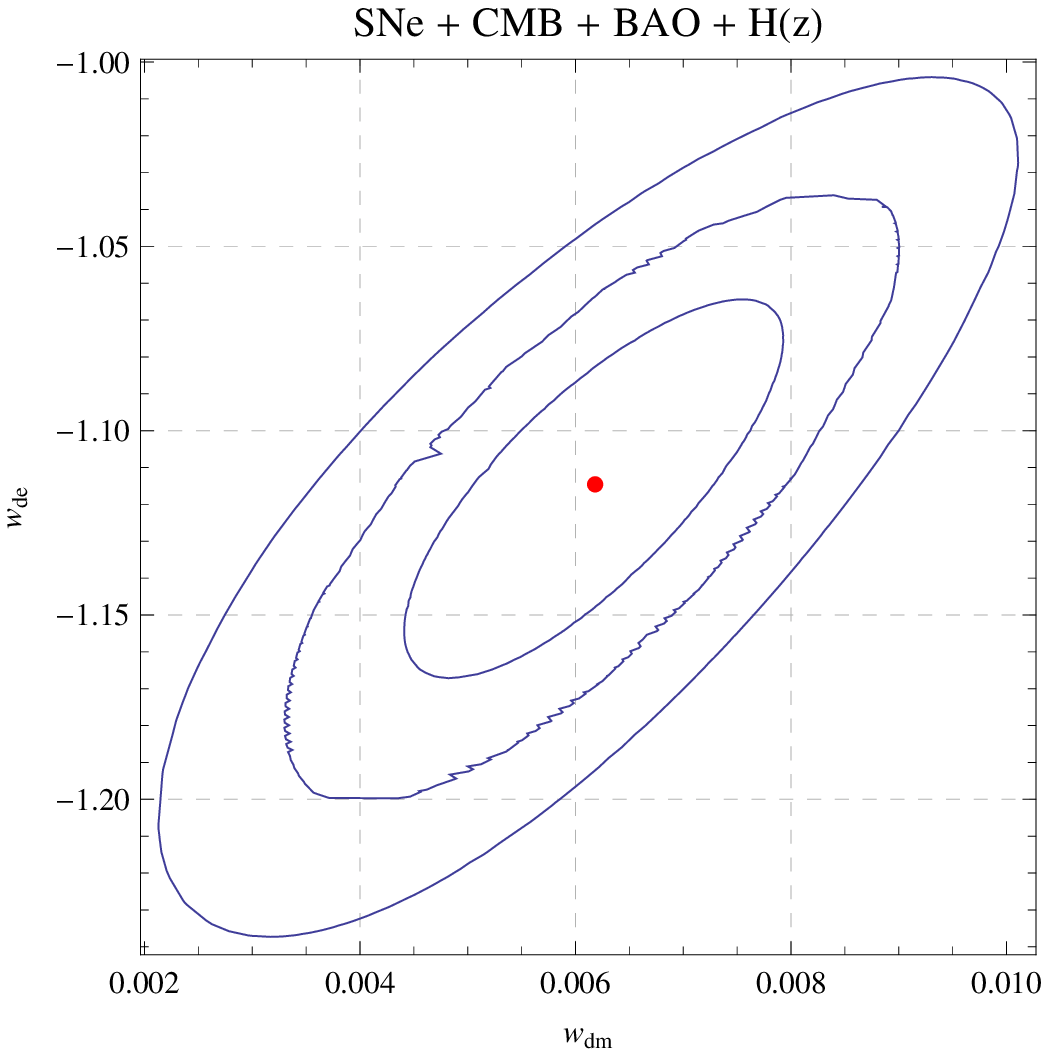}%
\hspace*{\fill}

\caption{Confidence intervals for $(w_{\rm dm}, w_{\rm de})$.
The upper left panel corresponds to the use of the SNe Ia data set
release (2012) ``Union 2.1'' of the SCP \cite{Union2.1:Suzuki2011}, through
the minimization of the $\chi^2$ function
(\ref{Chi2FunctionSNe}). The upper right panel corresponds to the use of
Hubble parameter data at
different redshifts using the $\chi^2$ function (\ref{Chi2FunctionHz}).
The lower left panel corresponds to the use of three observational data
$(\mathcal{R}, l_A, z_*)$ given by WMAP-7y \cite{WMAP7yKomatsu2011},
through the $\chi^2$ function (\ref{Chi2FunctionCMB3p}).
 And the lower right panel corresponds to the use of the total $\chi^2$
function
(\ref{Chi2FunctionTotal}) that contains the four type of cosmological
observations together SNe + CMB + BAO + $H(z)$.
The best estimated values for $(w_{\rm dm}, w_{\rm de})$ of each
panel are indicated with the red point and the magnitudes are shown in
table \ref{TableWdmWde}.
It is assumed a spatially flat FRW universe and for the baryon, dark matter,
radiation parameter densities and the Hubble constant it was assumed the
values of $\Omega_{\rm
b0}=0.0458$,  $\Omega_{\rm dm0}=0.23$, $\Omega_{\rm r0}=0.0000766$
\cite{WMAP7yKomatsu2011,SDSS7yReid2010} and $H_0 = 73.8$
km/s$\cdot$Mpc \cite{Riess:2011yx} respectively.
The contour plots correspond to 68.3\% (1$\sigma$),
95.4\% (2$\sigma$) and 99.73\% (3$\sigma$) of confidence level.}
\label{PlotGroupWdmWde}
\end{center}
\end{figure}



\begin{figure}
\begin{center}
\hfill%
\includegraphics[width=7cm]{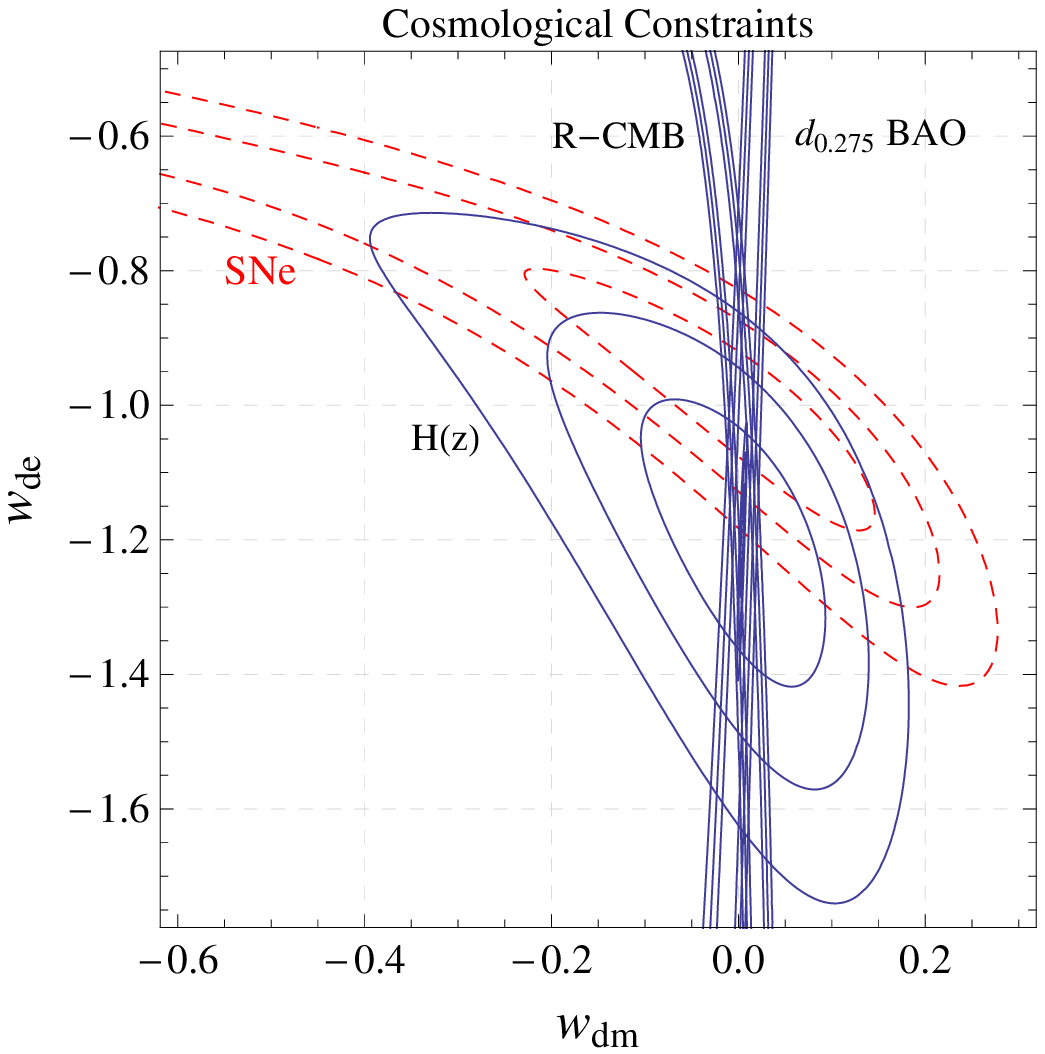}%
\hfill%
\includegraphics[width=7cm]{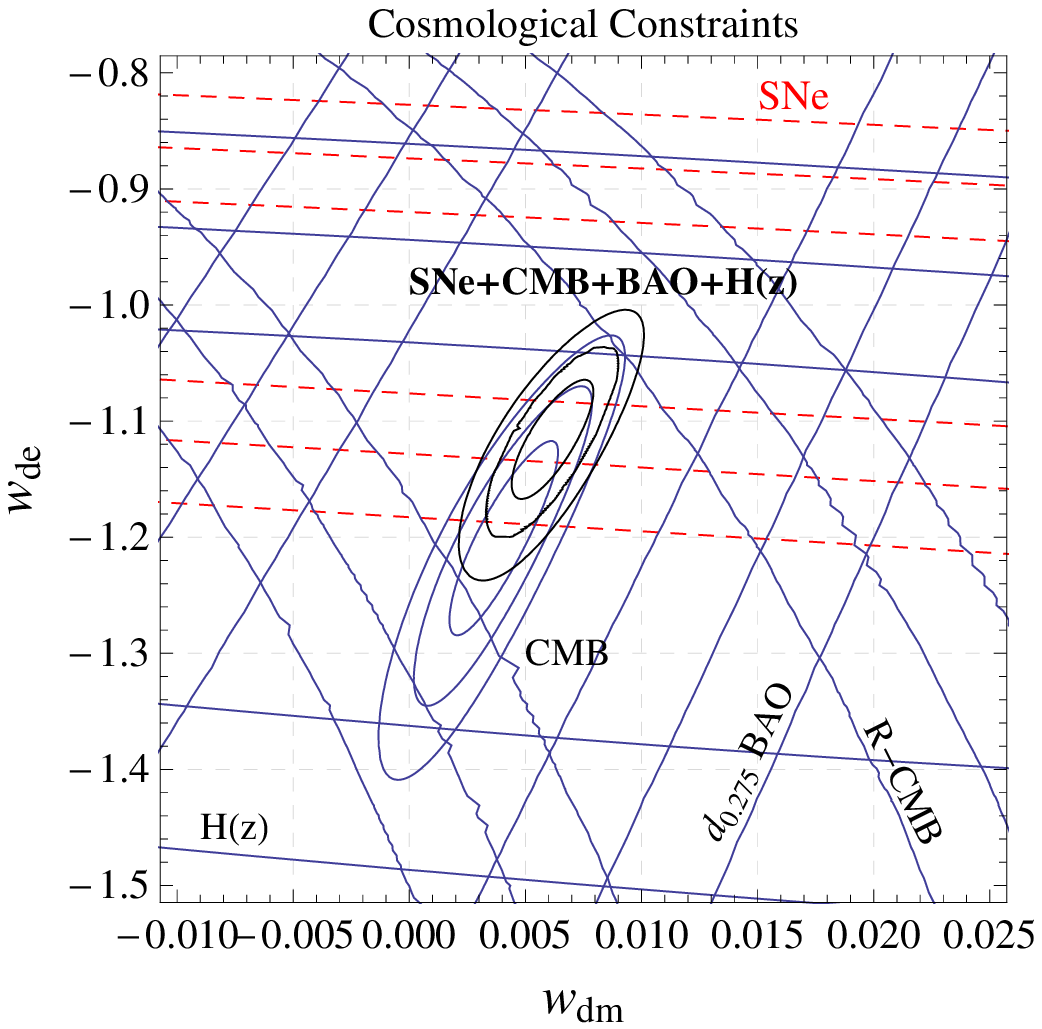}%
\hspace*{\fill}

\caption{Confidence intervals (CI) all together for $(w_{\rm dm}, w_{\rm
de})$ calculated with the different cosmological data sets (cf.
figure \ref{PlotGroupWdmWde} for CI separately for each cosmological
data set). 
The CI labeled by ``R-CMB'' and ``$d_{0.275}$ BAO''  come from the use of
the shift parameter $\mathcal{R}$ and the distance ratio $d_z$ at $z=0.275$ of
BAO computed through the
$\chi^2_{\rm}$ functions defined at (\ref{Chi2FunctionRCMB}) and
(\ref{Chi2FunctionBAO}) respectively.
The right panel corresponds to a zoom in of the left panel in order to
show the CI coming from the use of the 
$(\mathcal{R}, l_A, z_*)$
CMB distance priors and the joint SNe+CMB+BAO+$H(z)$. The best estimated
values are shown in table
\ref{TableWdmWde}.
The interval regions corresponds to 68.3\% (1$\sigma$),
95.4\% (2$\sigma$) and 99.73\% (3$\sigma$) of confidence level.}
\label{PlotGroupWdmWdeZoom}
\end{center}
\end{figure}



\begin{figure}
\begin{center}
\includegraphics[width=13cm]{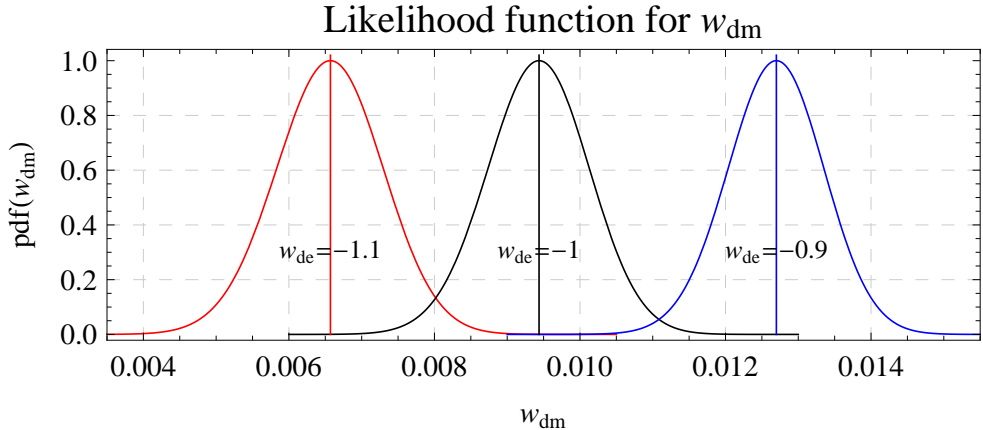}%
\caption{Likelihood functions for $w_{\rm dm}$  when it is assumed the values
of $w_{\rm de} = -1, -1.1, -0.9$ for the EoS of dark energy. See table
\ref{TableWdmAlone} for the best estimated values of $w_{\rm dm}$.
It is assumed a spatially flat FRW universe and for the baryon, dark matter,
radiation parameter densities and the Hubble constant it was assumed the
values of $\Omega_{\rm
b0}=0.0458$,  $\Omega_{\rm dm0}=0.23$, $\Omega_{\rm r0}=0.0000766$ and $H_0 =
73.8$ km/s$\cdot$Mpc respectively.}
\label{PlotGaussWdmAllTogetherJoinU21FDM}
\end{center}
\end{figure}



\begin{figure}
\begin{center}
\includegraphics[width=13cm]{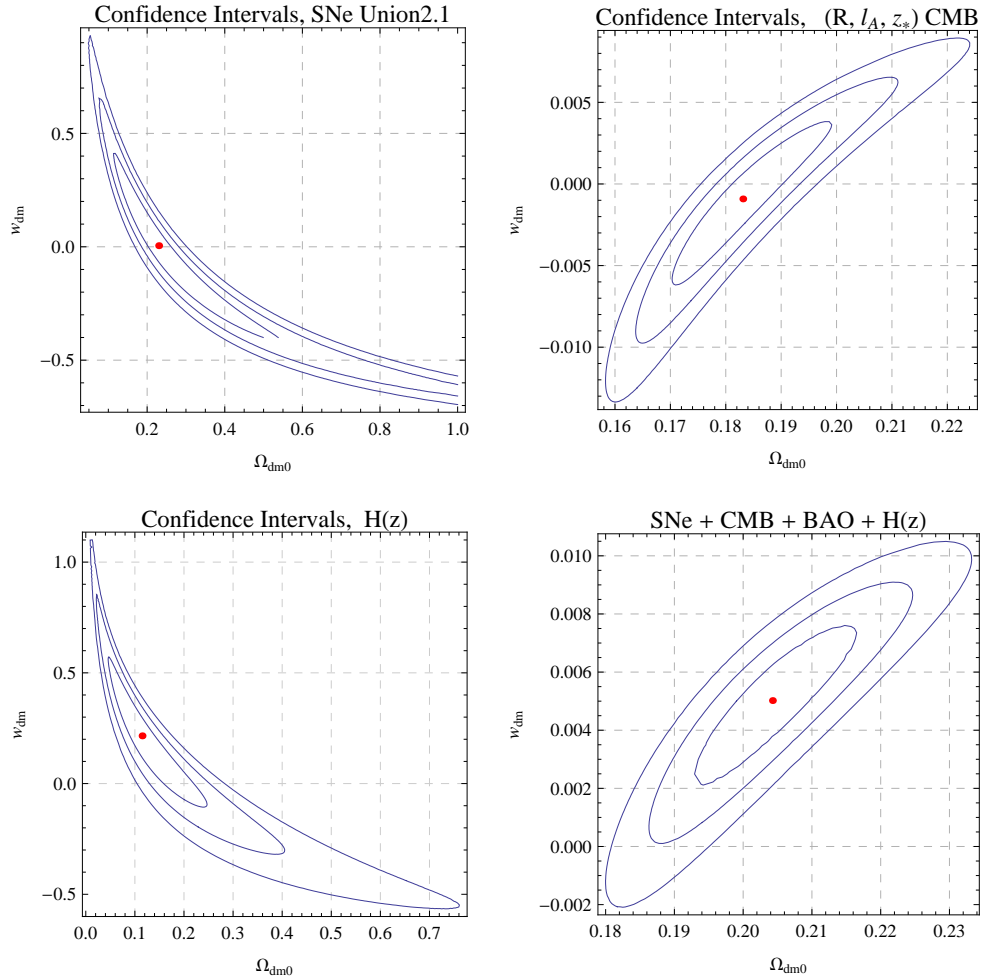}
\caption{Confidence intervals for the present-day value of the parameter
density of dark matter $\Omega_{\rm dm0}$ versus the $w_{\rm dm}$ of the
EoS of dark matter, and assuming $w_{\rm de} = -1$ for the dark energy.
The upper left, right, lower left and right panels correspond to the use of
SNe, $(\mathcal{R}, l_A, z_*)$ of CMB, $H(z)$ and the joint
SNe+CMB+BAO+$H(z)$ data sets respectively.
The best estimated values are indicated with the red point and the magnitudes
are shown in table \ref{TableWdmOdm}.
It is assumed a spatially flat FRW universe and for the baryon, dark matter,
radiation parameter densities and the Hubble constant it was assumed the
values of $\Omega_{\rm
b0}=0.0458$,  $\Omega_{\rm dm0}=0.23$, $\Omega_{\rm r0}=0.0000766$ and $H_0 =
73.8$ km/s$\cdot$Mpc respectively.
The interval regions correspond to 1, 2 and 3$\sigma$ of confidence level.}
\label{PlotsCIs}
\end{center}
\end{figure}


      \subsubsection{Cosmic Microwave Background Radiation}

We use the WMAP 7-years distance priors shown in table 9 of
\cite{WMAP7yKomatsu2011}, composed of the shift parameter $\mathcal{R}$,
the acoustic scale $l_A$ and the redshift of decoupling $z_*$.

The shift parameter $\mathcal{R}$ is defined as
\begin{equation}
\mathcal{R} = \frac{ H_0 \sqrt{\Omega_{\rm m0}}}{c}
(1+z_*) D_A(z_*)
\end{equation}
\noindent where $\Omega_{\rm m0}$ corresponds to the total present
pressureless matter in the Universe, i.e. $\Omega_{\rm m0} = \Omega_{\rm
b0} + \Omega_{\rm dm0}$, and $D_A$ is the
proper angular diameter distance given by
\begin{equation}
D_A(z)= \frac{c}{(1+z)H_0} \int^z_0 \frac{dz'}{E(z',w_{\rm dm})}.
\end{equation}
\noindent for a spatially flat Universe. With $\mathcal{R}$ we can defined
a $\chi^2$
function as
\begin{equation}\label{Chi2FunctionRCMB}
\chi^2_{\rm R-CMB}(w_{\rm dm}, H_0) \equiv \left(
\frac{\mathcal{R} - \mathcal{R}_{\rm obs}}{\sigma_{\mathcal{R}}}
\right)^2
\end{equation}
\noindent where $\mathcal{R}_{\rm obs} = 1.725$ is the ``observed'' value
of the shift parameter and $\sigma_{\mathcal{R}}=0.018$ the standard deviation of
the measurement (cf. table 9 of \cite{WMAP7yKomatsu2011}).

The acoustic scale $l_A$ is defined as
\begin{equation}
l_A \equiv (1+z_*)\frac{\pi D_A(z_*)}{r_s(z_*)},
\end{equation}
\noindent where $r_s(z_*)$ corresponds to the comoving sound horizon
at the decoupling epoch of photons, $z_*$, given by
\begin{equation}
r_s(z)=\frac{c}{\sqrt{3}} \int_0^{1/(1+z)} \frac{da}{a^2 H(a)
\sqrt{1+(3 \Omega_{\rm b0} / 4 \Omega_{\gamma 0})a}}
\end{equation}
\noindent where as mentioned above, we use $\Omega_{\gamma 0} = 2.469 \times
10^{-5} h^{-2}$ as the present-day photon energy density parameter, and
$\Omega_{b0} = 0.02255 h^{-2}$ as the baryonic
matter component, as reported by Komatsu et al. 2011
\cite{WMAP7yKomatsu2011}.
We compute the theoretical value of $z_*$ from the fitting formula proposed
by Hu and Sugiyama \cite{Hu:1995en}
\begin{equation}
z_*=1048 \left[ 1+0.00124 (\Omega_{\rm b0} h^2)^{-0.738} \right] \left[ 1
+ g_1 (\Omega_{\rm m0} h^2)^{g_2} \right],
\end{equation}
\noindent where
\begin{equation}
g_1 = \frac{0.0783
(\Omega_{\rm b0}h^2)^{-0.238}}{1+39.5(\Omega_{\rm b0}h^2)^{0.763}}, \; \;
\; \; \; g_2= \frac{0.560}{1+21.1(\Omega_{\rm b0} h^2)^{1.81}}.
\end{equation}

The $\chi^2$ function using the three distance priors $(l_A, \mathcal{R},
z_*)$ is
defined as
\begin{equation}\label{Chi2FunctionCMB3p}
\chi^2_{\rm CMB}(w_{\rm dm}, H_0) =  \sum_{i,j=1}^3 (x_i
- d_i)(C^{-1})_{ij}(x_j - d_j)
\end{equation}
\noindent where $x_i \equiv (l_A, \mathcal{R}, z_*) $ are the theoretical
values
predicted
by the model and $d_i \equiv (l_A = 302.09, \mathcal{R}=1.725, z_*=1091.3)$
are the observed ones. For $H_0$  it was assumed the latest reported
value of $H_0 = 73.8$ km/s$\cdot$Mpc \cite{Riess:2011yx}.
The $C^{-1}_{ij}$ is the inverse covariance
matrix with entries \cite{WMAP7yKomatsu2011}


\begin{equation}
C^{-1} =
\begin{pmatrix}
2.305 & 29.698 & -1.333 \\
29.698 & 6825.27 & -113.180 \\
1.333 & 113.180 & 3.414
\end{pmatrix}
\end{equation}


\begin{figure}
\begin{center}
\includegraphics[width=14cm]{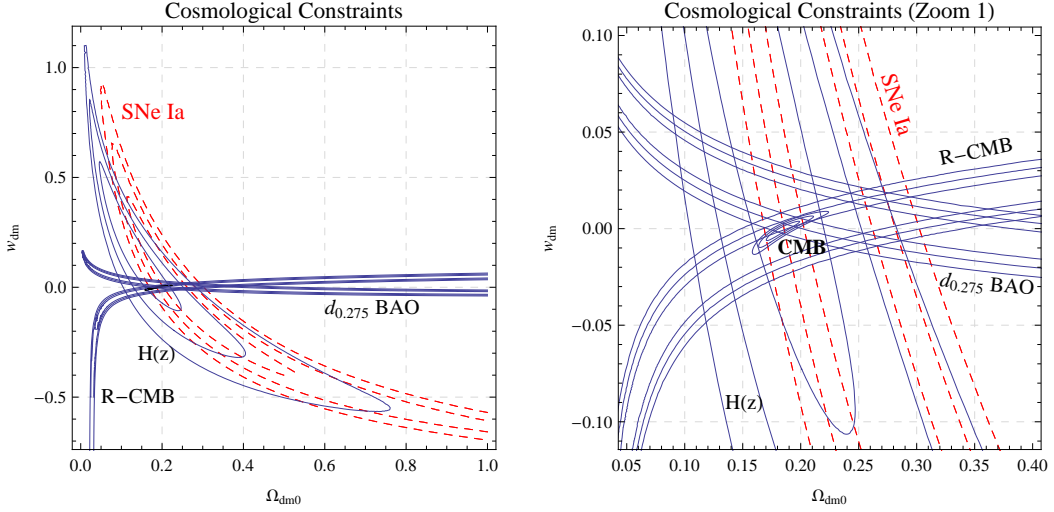}
\caption{Confidence intervals (CI) for $(\Omega_{\rm dm0},
w_{\rm dm})$, calculated with the different cosmological data sets (see
figure \ref{PlotsCIs}). It is assumed a value of $w_{\rm de} = -1$ for
the parameter of EoS of dark energy.
The right panel corresponds to a ``zoom in'' of the left panel in order to
show the tiny CI that come from the use of the $(\mathcal{R}, l_A, z_*)$
CMB distance priors through the $\chi^2_{\rm CMB}$ function
(\ref{Chi2FunctionCMB3p}) and labeled as ``CMB''.
The CI from the total $\chi^2$ function (\ref{Chi2FunctionTotal}) are even
smaller than those of the CMB and so they are shown in figure
\ref{PlotsAllTogetherZoom}. The best estimated values are shown in table
\ref{TableWdmOdm}.
The interval regions corresponds to 68.3\% (1$\sigma$),
95.4\% (2$\sigma$) and 99.73\% (3$\sigma$) of confidence level.}
\label{PlotsAllTogether}
\end{center}
\end{figure}

\begin{figure}
\begin{center}
\includegraphics[width=10cm]{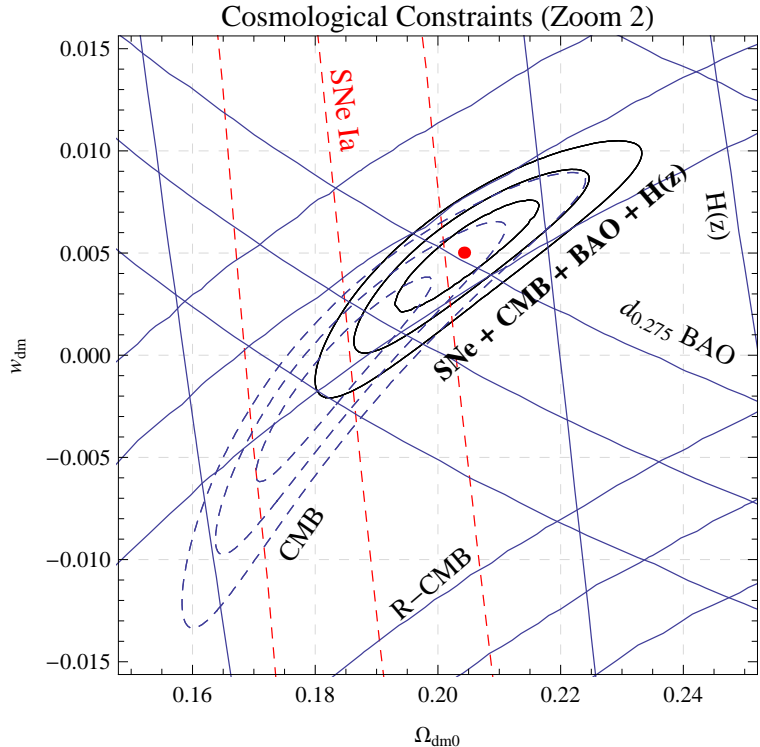}
\caption{Confidence intervals for $(\Omega_{\rm dm0}, w_{\rm
dm})$. This figure corresponds to a ``zoom in'' of the figures
\ref{PlotsCIs} and \ref{PlotsAllTogether} to show the CI that come from
the use of all the observational data sets together through the
total $\chi^2$ function defined in (\ref{Chi2FunctionTotal}) and labeled as
``SNe + CMB + BAO + $H(z)$''. The best estimated
values computed with the total $\chi^2$ function are $w_{\rm
dm}=0.005$, $\Omega_{\rm dm0} = 0.204$ (see table \ref{TableWdmOdm}).
The intervals regions correspond to 68.3\% (1$\sigma$), 95.4\% (2$\sigma$)
and 99.73\% (3$\sigma$) of confidence level.
Notice that $w_{\rm dm}>0$ with 95\% confidence level, suggesting a
\textit{warm} dark matter. }
\label{PlotsAllTogetherZoom}
\end{center}
\end{figure}



\begin{figure}
\begin{center}
\hfill%
\includegraphics[width=7cm]{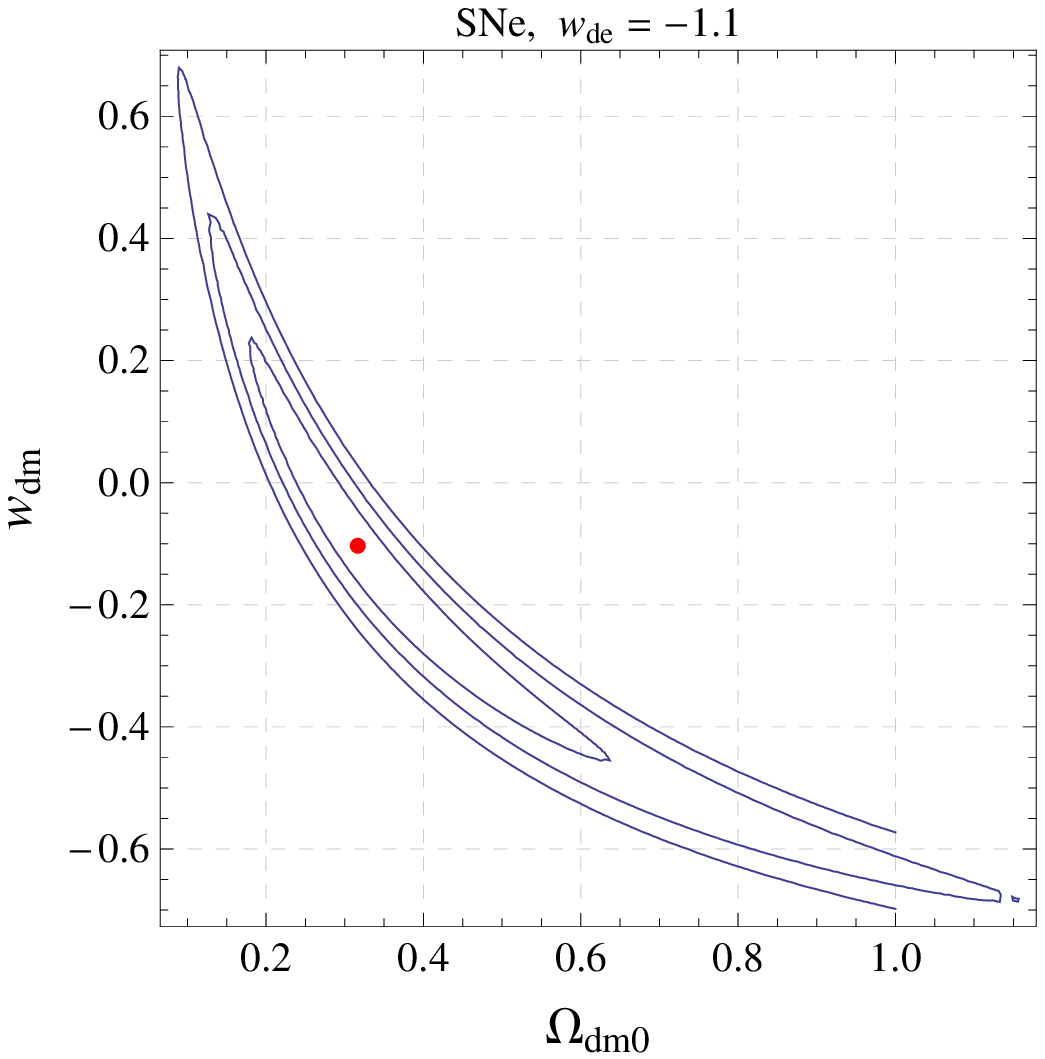}%
\hfill%
\includegraphics[width=7cm]{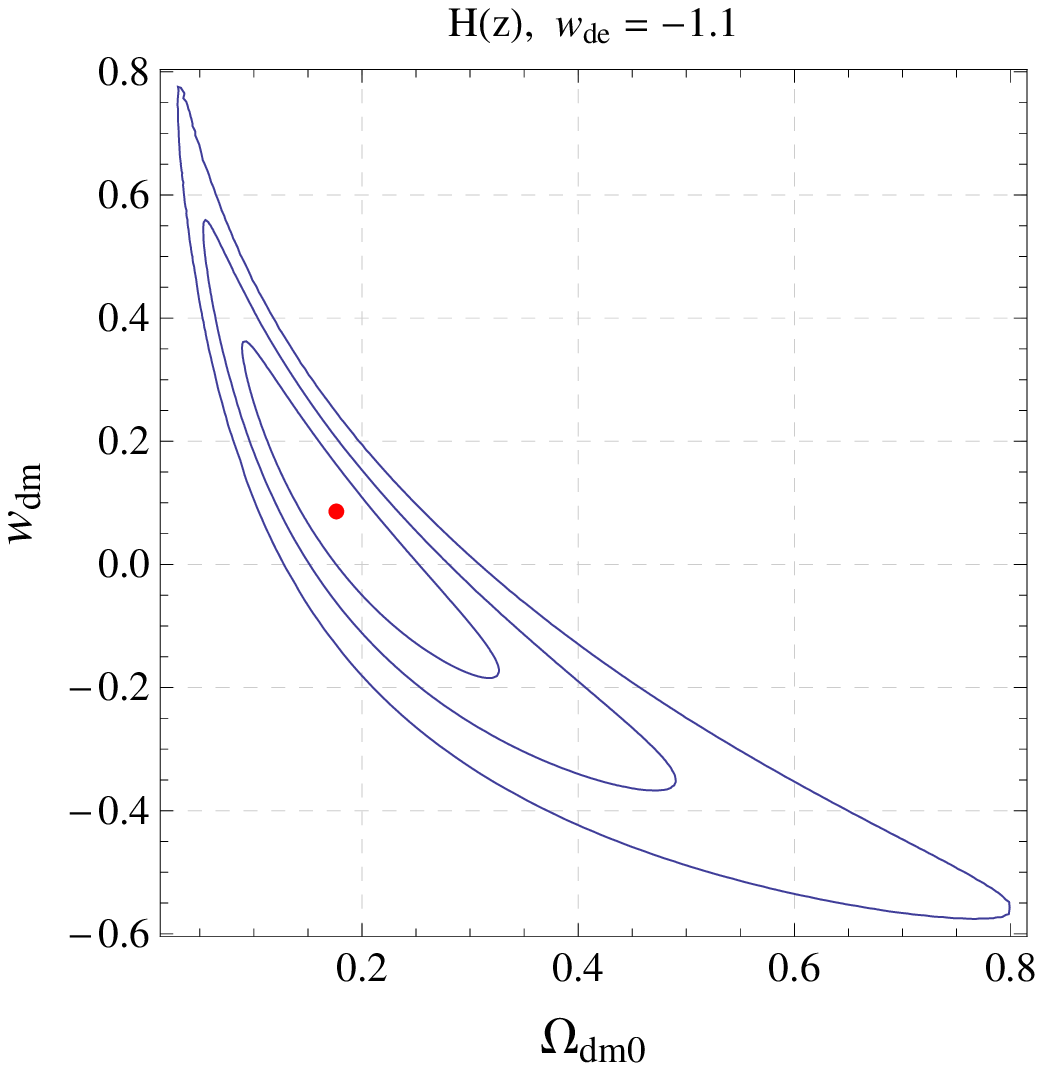}%
\hspace*{\fill}

\hfill%
\includegraphics[width=7cm]{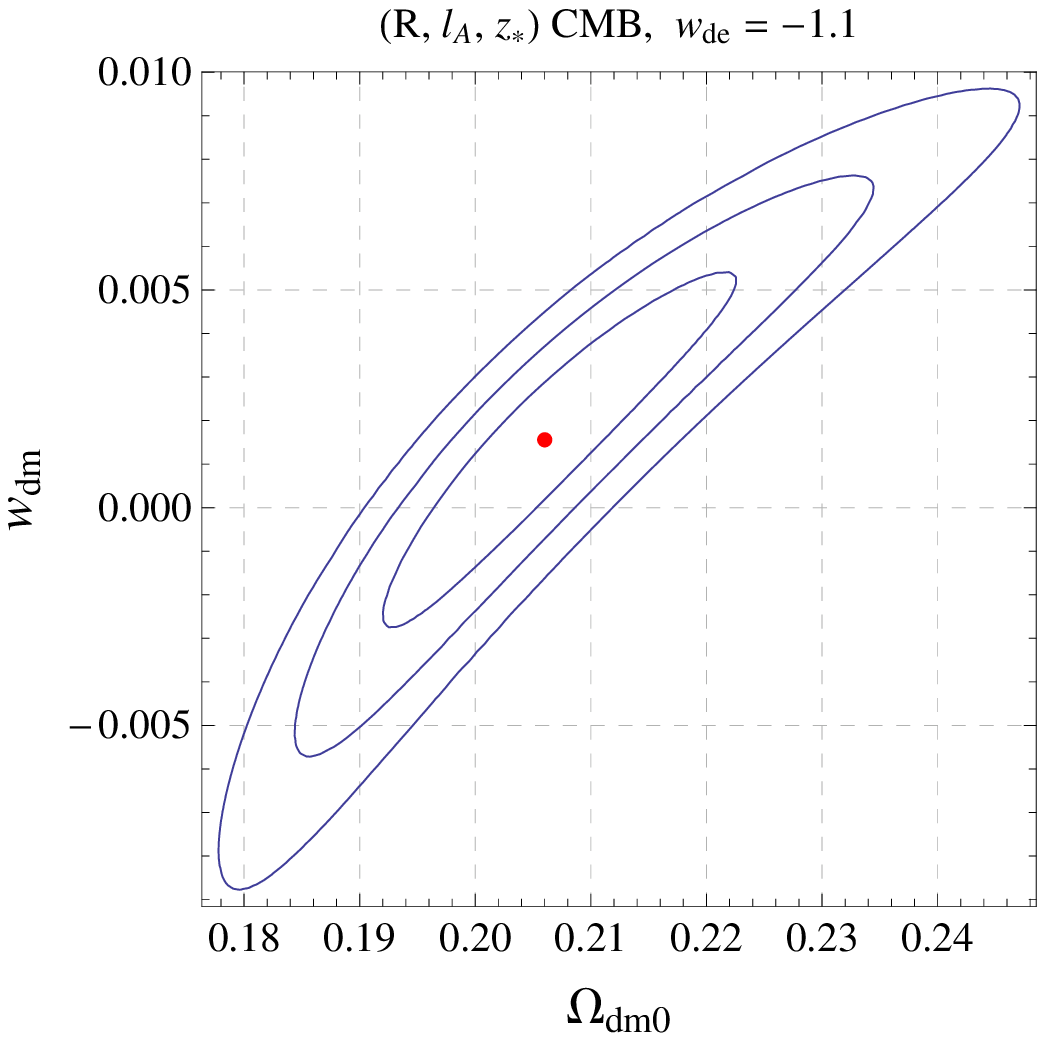}%
\hfill%
\includegraphics[width=7cm]{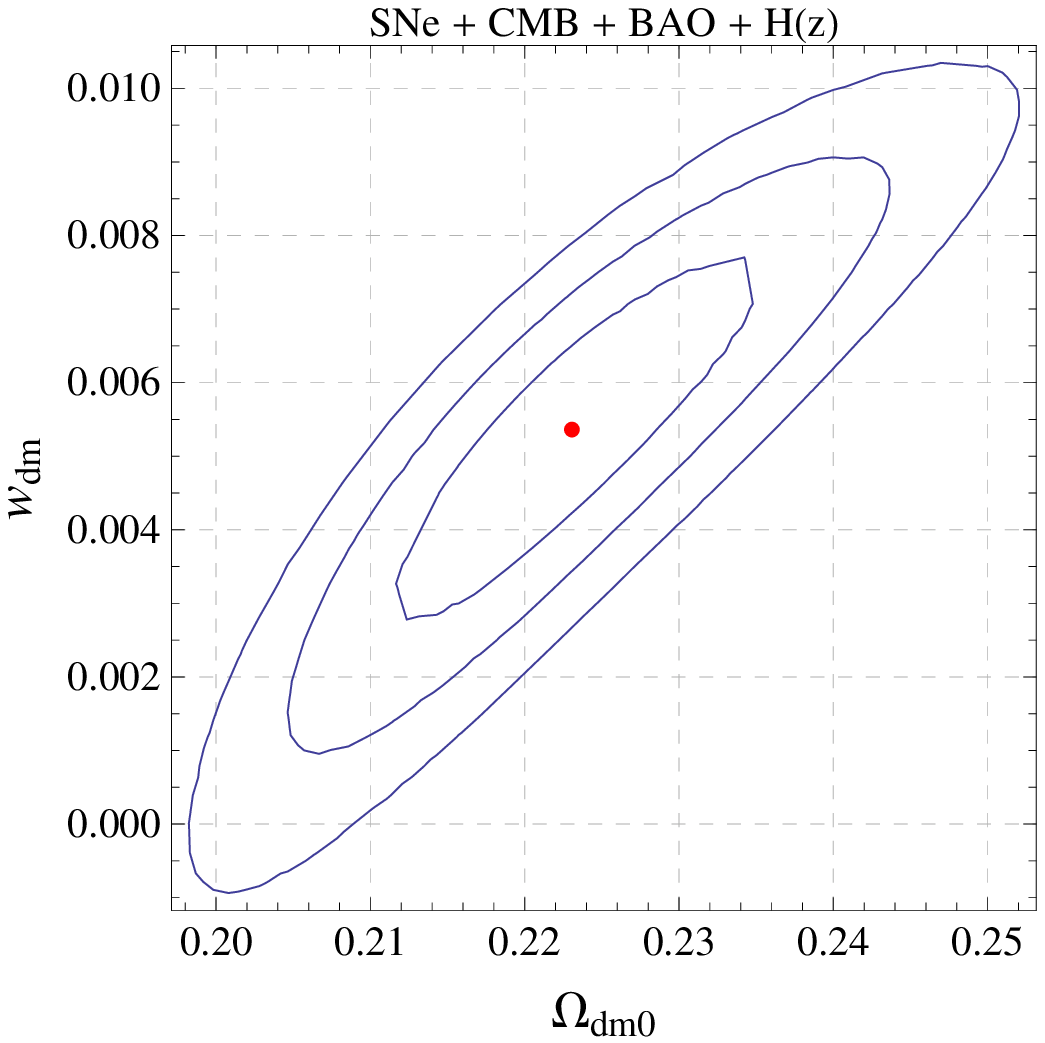}%
\hspace*{\fill}
\caption{Confidence intervals for $(\Omega_{\rm dm0}, w_{\rm
dm})$ when it is assumed the value of $w_{\rm de} = -1.1$ for the parameter of
EoS of dark energy, i.e., a phantom dark energy. See table \ref{TableWdmOdm}
for the values of the best estimates. The interval regions corresponds to
68.3\% (1$\sigma$), 95.4\% (2$\sigma$) and 99.73\% (3$\sigma$) of confidence
level.}
\label{PlotsOdmWdmWdeM11}
\end{center}
\end{figure}



\begin{figure}
\begin{center}
\hfill%
\includegraphics[width=5cm]{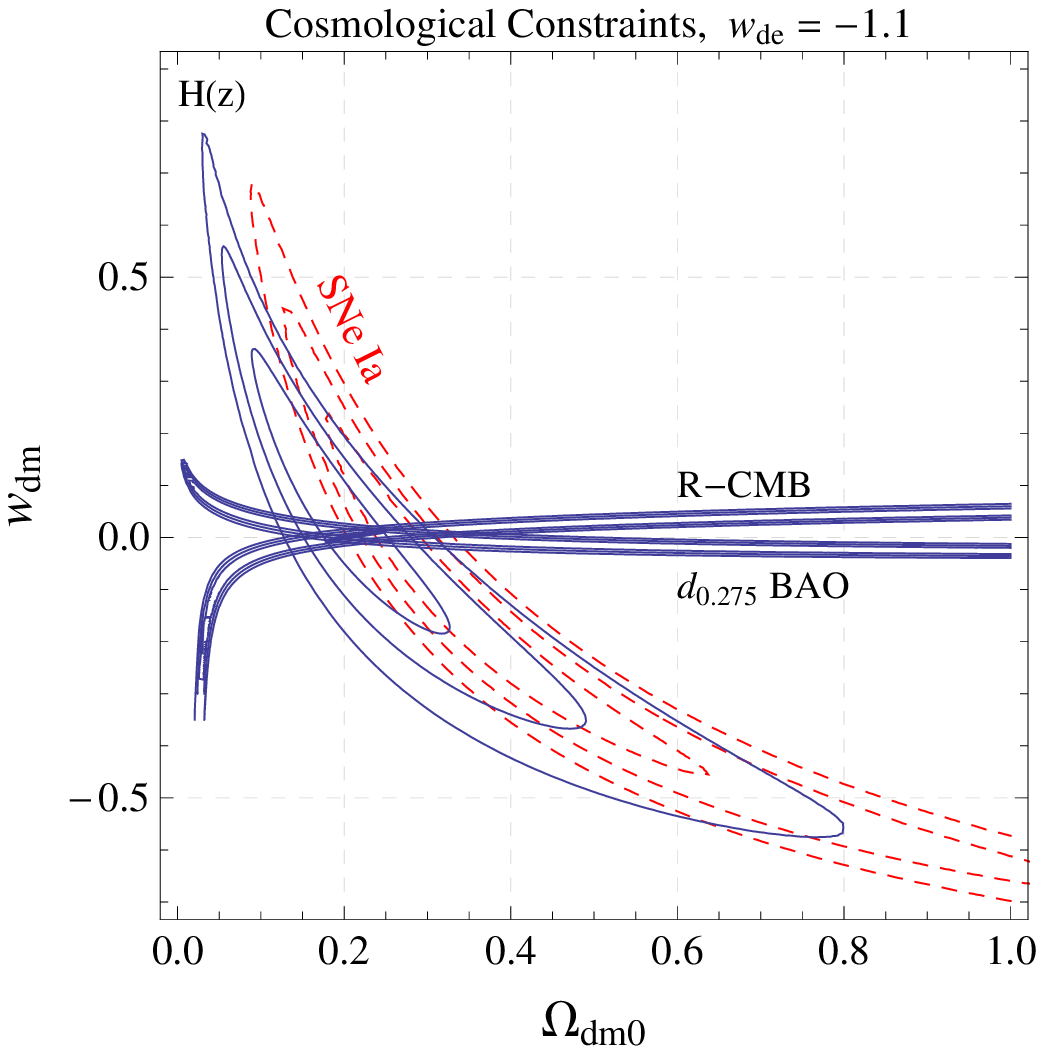}%
\hfill%
\includegraphics[width=5cm]{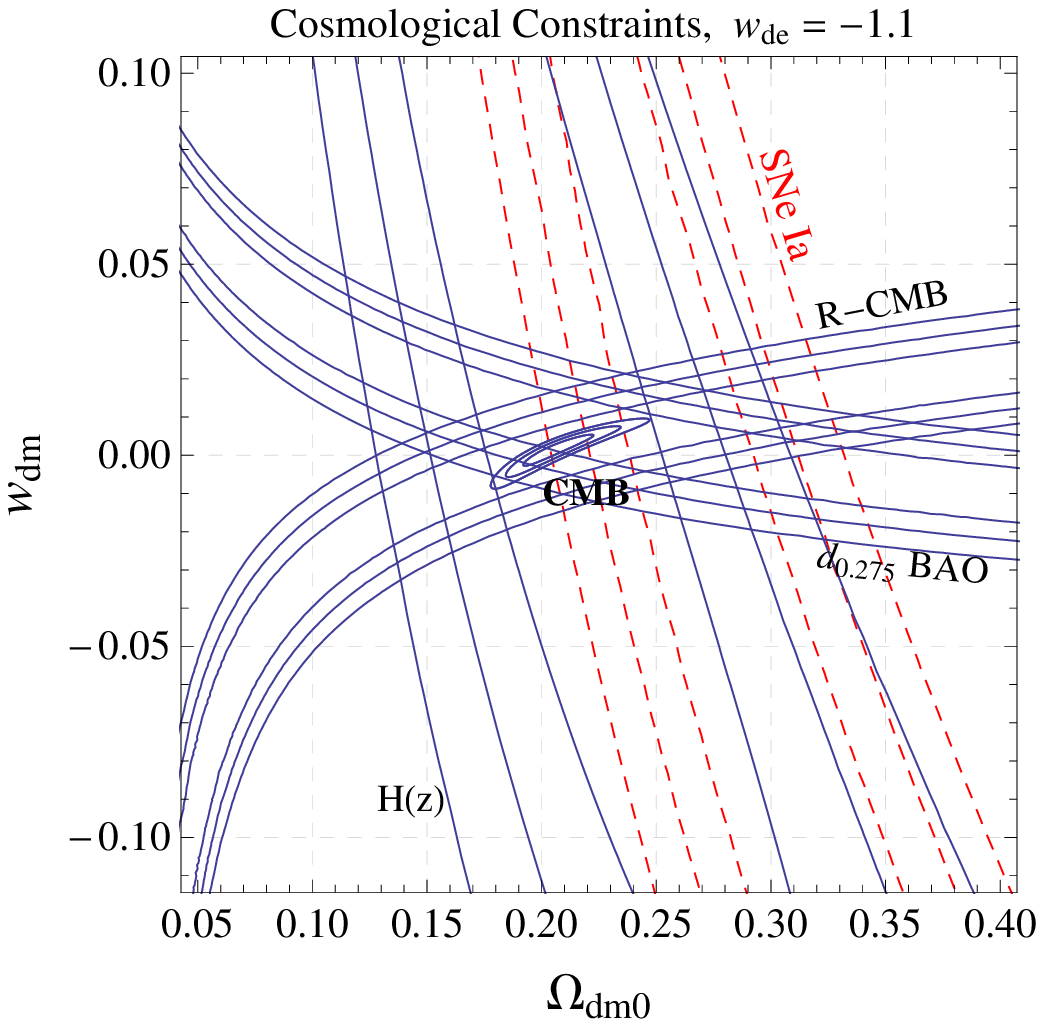}%
\hfill%
\includegraphics[width=5cm]{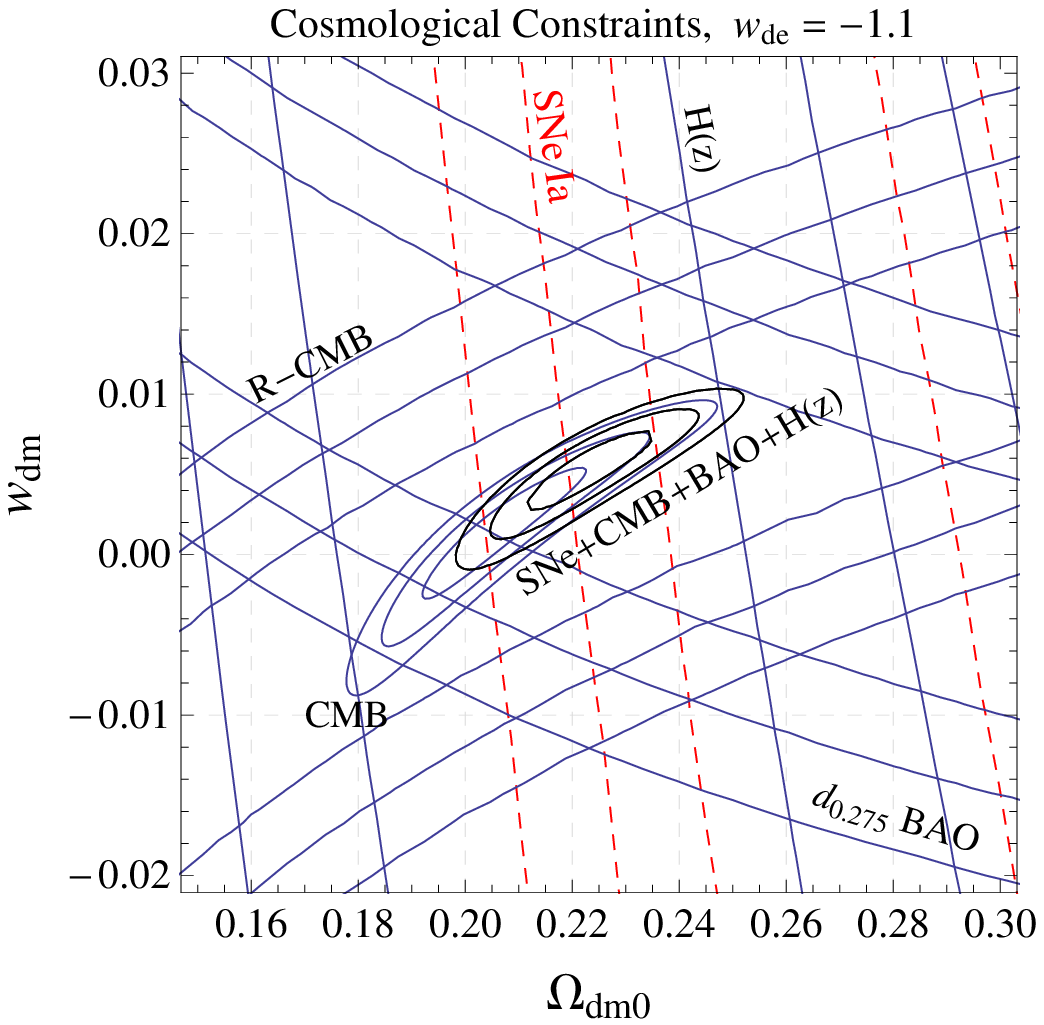}%
\hspace*{\fill}

\caption{Confidence intervals for $(\Omega_{\rm dm0}, w_{\rm
dm})$ when it is assumed the value of $w_{\rm de} = -1.1$ for the parameter of
EoS of dark energy, i.e., a phantom dark energy. See also figure
\ref{PlotsOdmWdmWdeM11}. Table \ref{TableWdmOdm} shows the values of the best
estimates for this case. The central and right panels correspond to a zoom in
of the left panel. 
The interval regions corresponds to 68.3\%
(1$\sigma$), 95.4\% (2$\sigma$) and 99.73\% (3$\sigma$) of confidence level.}
\label{PlotsOdmWdmWdeM11Zoom}
\end{center}
\end{figure}



\begin{figure}
\begin{center}
\hfill%
\includegraphics[width=7cm]{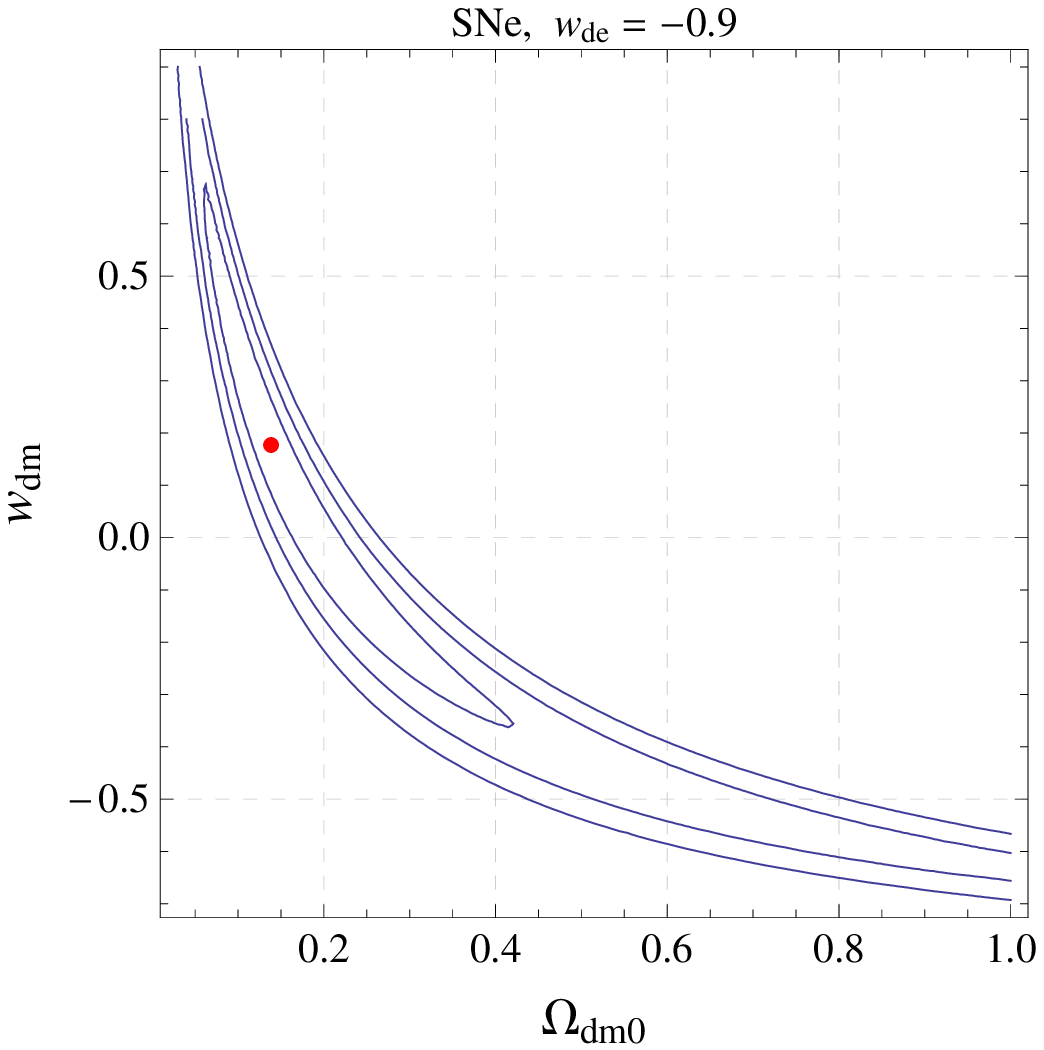}%
\hfill%
\includegraphics[width=7cm]{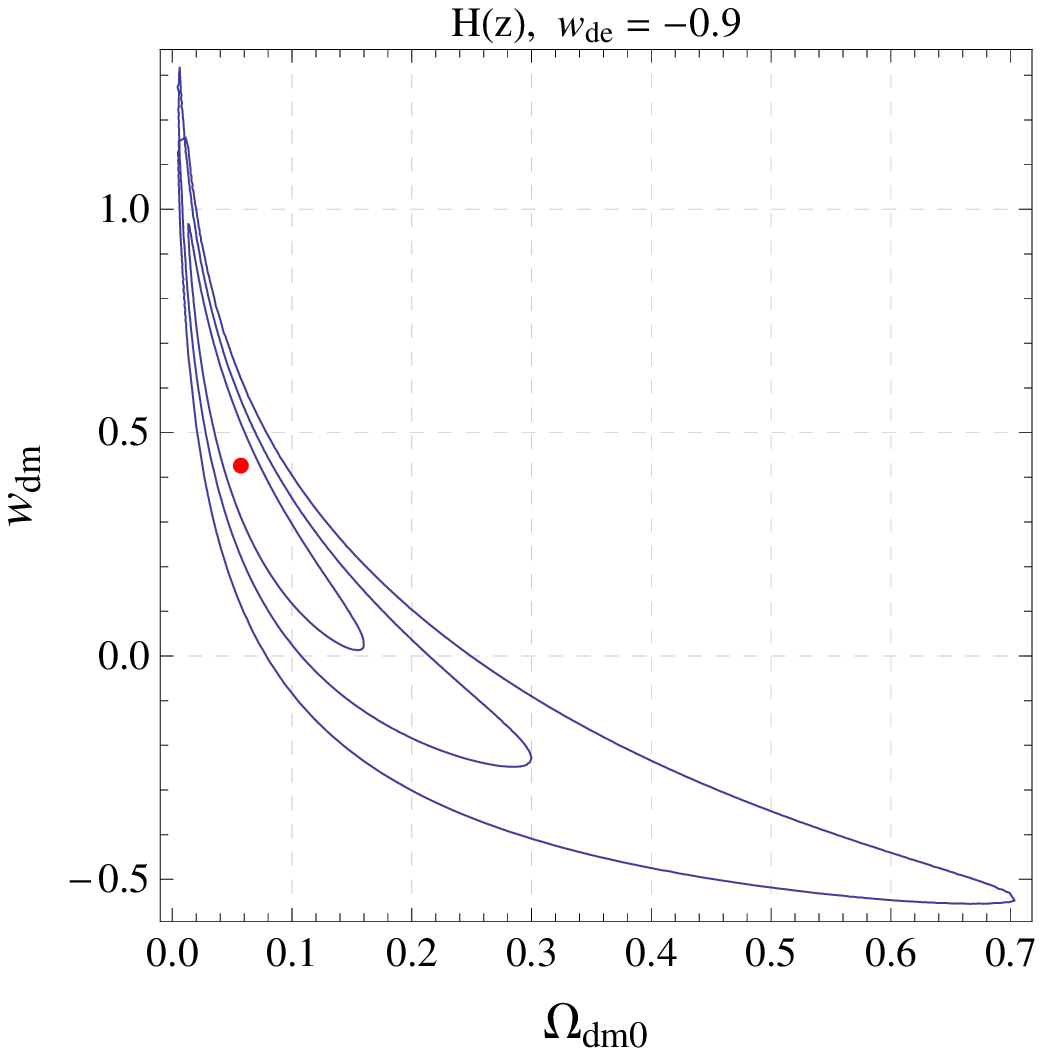}%
\hspace*{\fill}

\hfill%
\includegraphics[width=7cm]{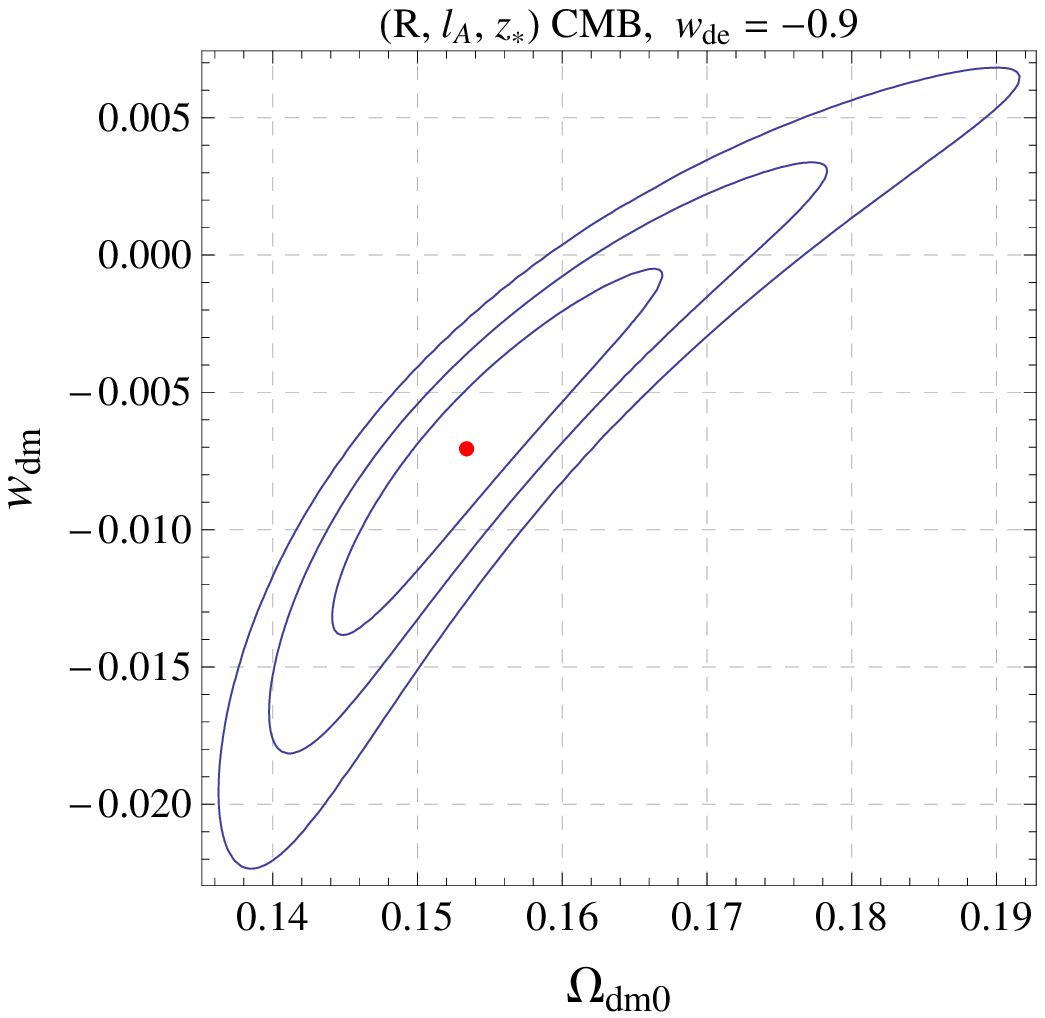}%
\hfill%
\includegraphics[width=7cm]{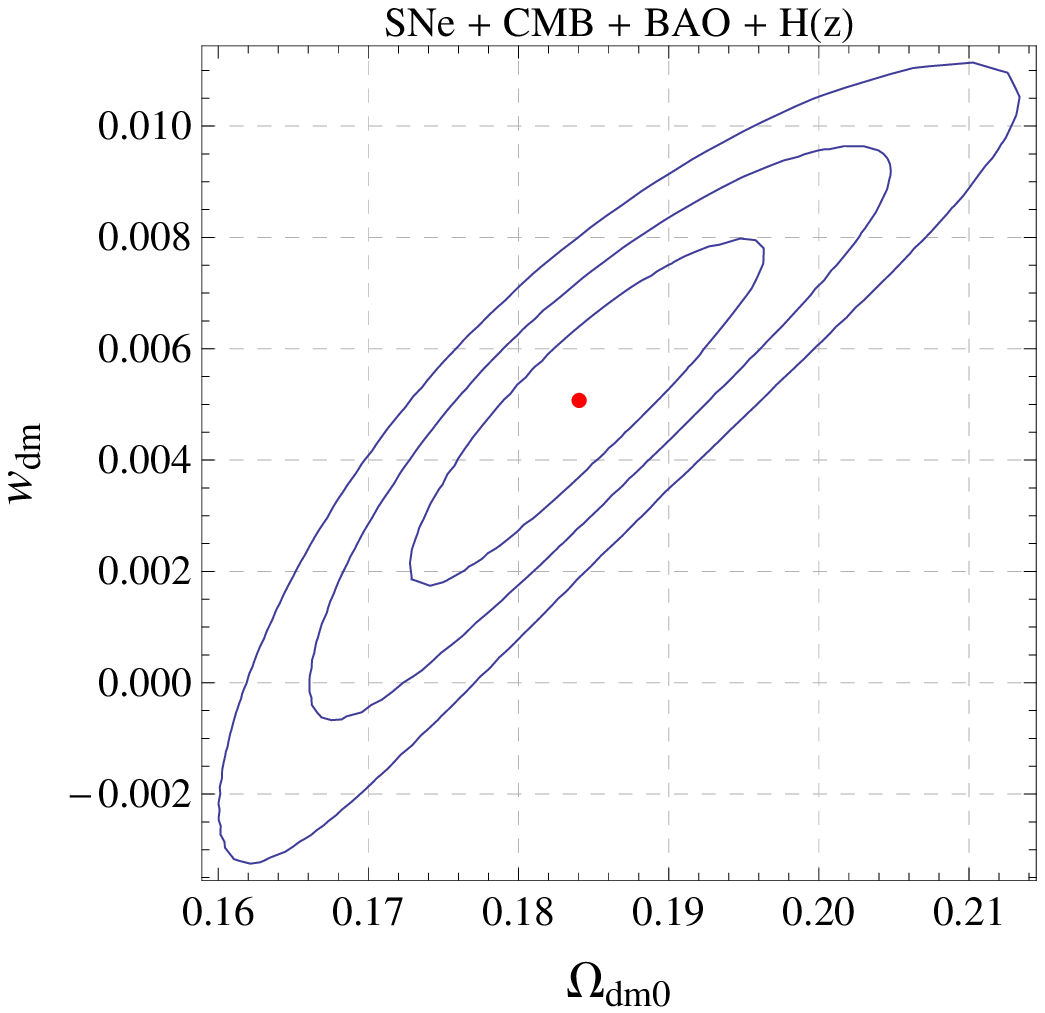}%
\hspace*{\fill}
\caption{Confidence intervals for $(\Omega_{\rm dm0}, w_{\rm
dm})$ when it is assumed the value of $w_{\rm de} = -0.9$ for the parameter of
EoS of dark energy. See table \ref{TableWdmOdm} for the values of
the best estimates. The interval regions corresponds to 68.3\% (1$\sigma$),
95.4\% (2$\sigma$) and 99.73\% (3$\sigma$) of confidence level.} 
\label{PlotsOdmWdmWdeM09}
\end{center}
\end{figure}



\begin{figure}
\begin{center}
\hfill%
\includegraphics[width=5cm]{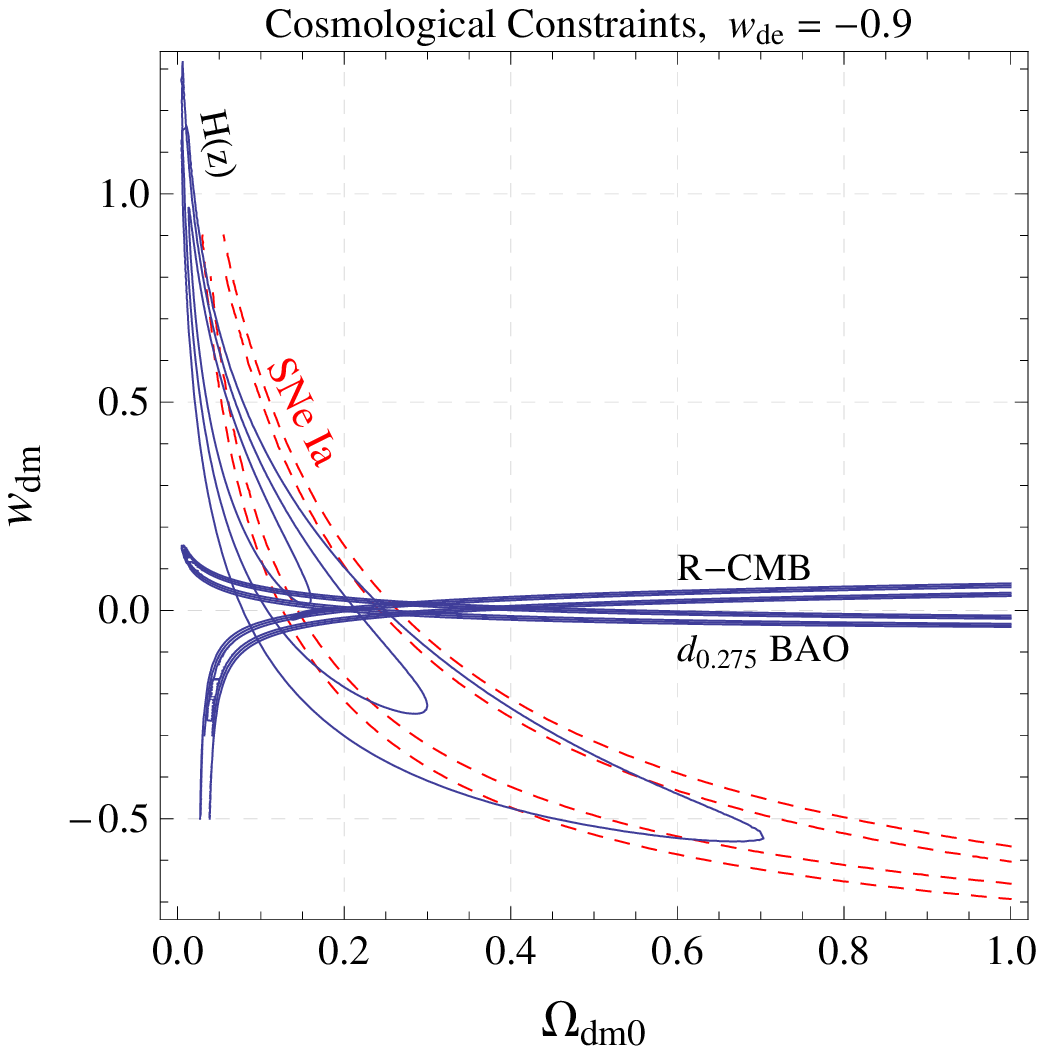}%
\hfill%
\includegraphics[width=5cm]{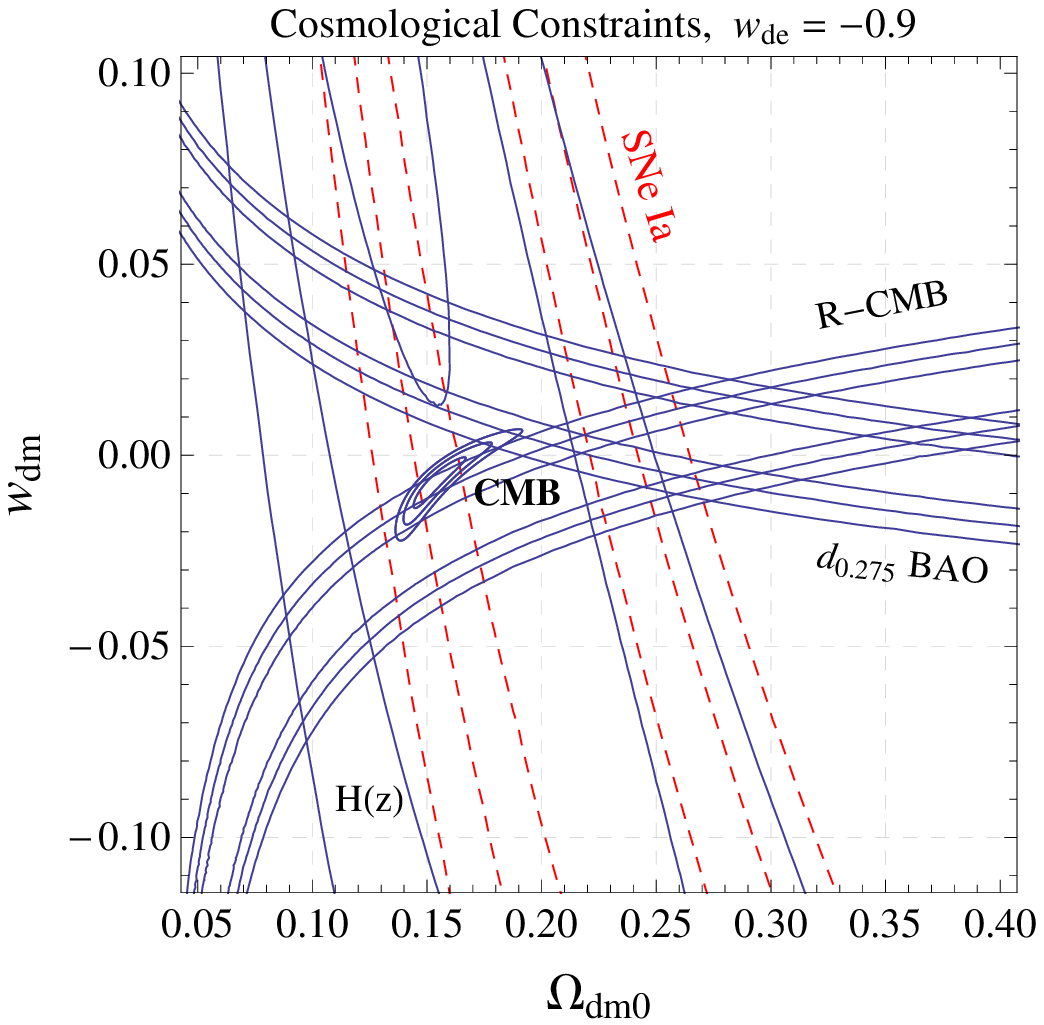}%
\hfill%
\includegraphics[width=5cm]{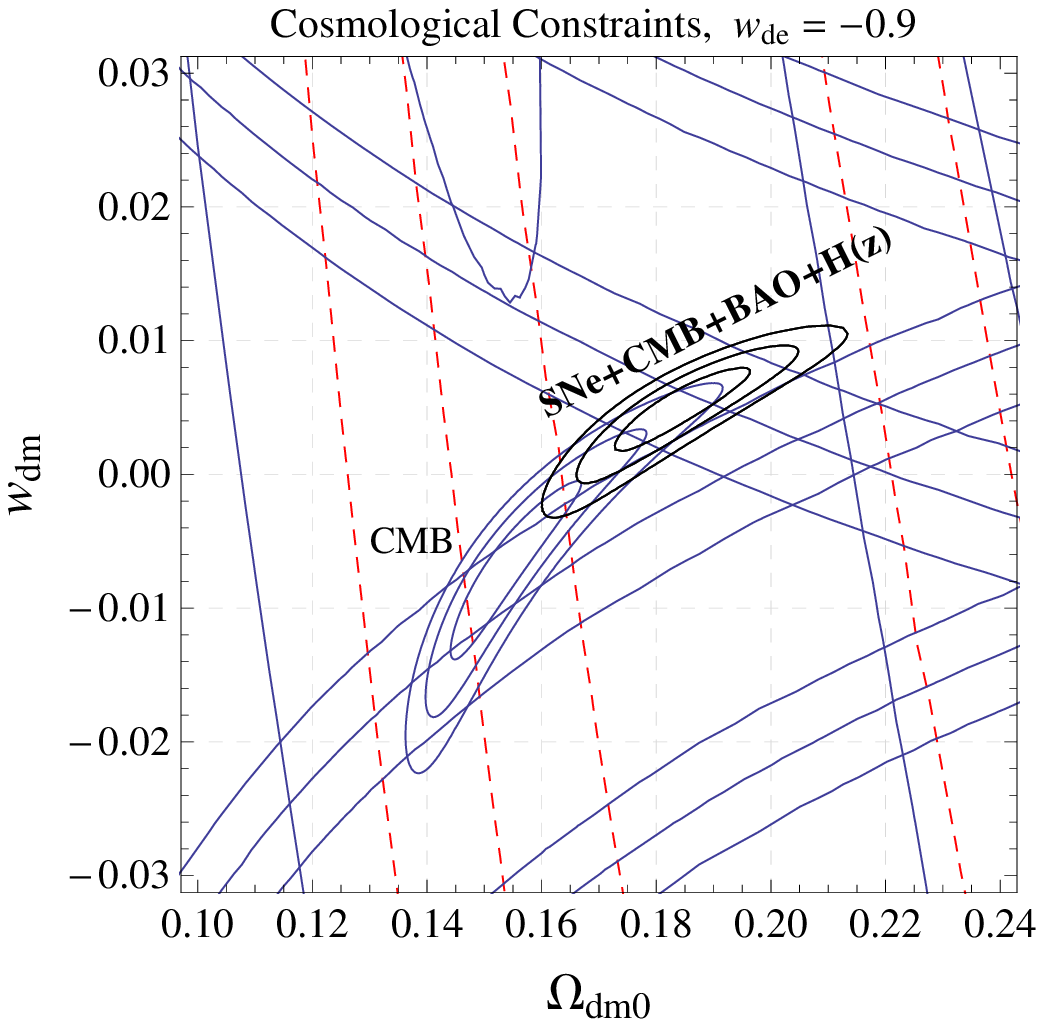}%
\hspace*{\fill}

\caption{Confidence intervals for $(\Omega_{\rm dm0}, w_{\rm dm})$ when it is
assumed the value of $w_{\rm de} = -0.9$ for the parameter of
EoS of dark energy. See also figure \ref{PlotsOdmWdmWdeM09}. Table
\ref{TableWdmOdm} shows the values of the best estimates for this case. The
central and right panels correspond to a zoom in
of the left one. The interval regions corresponds to 68.3\%
(1$\sigma$), 95.4\% (2$\sigma$) and 99.73\% (3$\sigma$) of confidence level.} 
\label{PlotsOdmWdmWdeM09Zoom}
\end{center}
\end{figure}


      \subsubsection{Baryon Acoustic Oscillations}

We use the baryon acoustic oscillation (BAO) data from the SDSS
7-years release \cite{Percival:2009xn}, expressed in terms of the distance
ratio $d_z$ at $z=0.275$ defined as
\begin{equation}
d_{0.275} \equiv \frac{r_s(z_d)}{D_V(0.275)}
\end{equation}
\noindent where $z_d$ is the redshift at the baryon drag epoch
computed from the fitting formula \cite{Eisenstein:1997ik}
\begin{align}
z_d &= 1291 \frac{(\Omega_{\rm m0} h^2)^{0.251}}{1+0.659(\Omega_{\rm m0}
h^2)^{0.828}} \left[ 1 + b_1 (\Omega_{\rm m0} h^2)^{b_2} \right], \\
b_1 &= 0.313 (\Omega_{\rm m0} h^2)^{-0.419} \left[1 + 0.607 (\Omega_{\rm
m0} h^2)^{0.674} \right], \\
b_2 &= 0.238 (\Omega_{\rm m0} h^2)^{0.223}.
\end{align}

For a flat Universe, $D_V(z)$ is defined as

\begin{equation}
D_V(z) = c \left[ \left( \int_0^z \frac{dz'}{H(z')} \right)^2
\frac{z}{H(z)} \right]^{1/3}.
\end{equation}
\noindent It contains the information of the visual distortion of a
spherical object due the non-Euclidianity of the FRW spacetime.

The value $d_{0.275}^{\rm obs}$ contains the information of the other two pivots,
$d_{0.2}$ and $d_{0.35}$, usually used by other authors, with a
precision of $0.04\%$ \cite{Percival:2009xn}.

The $\chi^2$ function for BAO is defined as
\begin{equation}\label{Chi2FunctionBAO}
\chi^2_{\rm BAO}( w_{\rm dm}, H_0) \equiv \left(
\frac{d_{0.275} - d_{0.275}^{\rm
obs}}{\sigma_{d}}
\right)^2
\end{equation}
\noindent where $d_{0.275}^{\rm obs} = 0.139$ is the observed
value and $\sigma_{d} = 0.0037$ the standard deviation
of the measurement \cite{Percival:2009xn}. For $H_0$  it was assumed the
latest reported value of $H_0 = 73.8$ km/s$\cdot$Mpc \cite{Riess:2011yx}.

      \subsubsection{Hubble expansion rate}
\label{SectionHz}

For the Hubble parameter, we use the 13 available data, 11 data come from
the table 2 of Stern et al. (2010) \cite{Stern:2009ep} and the two following data
come from Gaztanaga et al. 2010
\cite{Gaztanaga:2008xz}: $H(z=0.24)=79.69 \pm 2.32$ and $H(z=0.43)=
86.45 \pm 3.27$ km/s/Mpc.
For the present value of the Hubble parameter, we take the value reported
by Riess et al 2011 \cite{Riess:2011yx}: $H(z=0) \equiv H_0
= 73.8 \pm 2.4$ km/s/Mpc.
The $\chi^2$ function is defined as
\begin{equation}\label{Chi2FunctionHz}
\chi^2_{\rm H}( w_{\rm dm},  H_0) = \sum_i^{13} \left(
\frac{H(z_i,  w_{\rm dm}) - H_i^{\rm obs}}{\sigma_{H} }
\right)^2
\end{equation}

\noindent where $H(z_i)$ is the theoretical value predicted by the
model and $H_i^{\rm obs}$ is the observed value with its standard deviation
$\sigma_{H}$.

Finally, with the $\chi^2$ functions defined above we construct the total
$\chi^2$ function given by
\begin{equation}\label{Chi2FunctionTotal}
\chi^2 = \chi^2_{\rm SNe} + \chi^2_{\rm CMB} + \chi^2_{\rm BAO} +
\chi^2_{\rm H}.
\end{equation}

We minimize this function with respect to the set of parameters $(w_{\rm dm},
w_{\rm de})$, $(w_{\rm dm}, \Omega_{\rm dm0})$ and $w_{\rm dm}$ alone, to
compute their best estimated values and confidence intervals or likelihood
functions.



\begin{figure}
\begin{center}
\hfill%
\includegraphics[width=7cm]{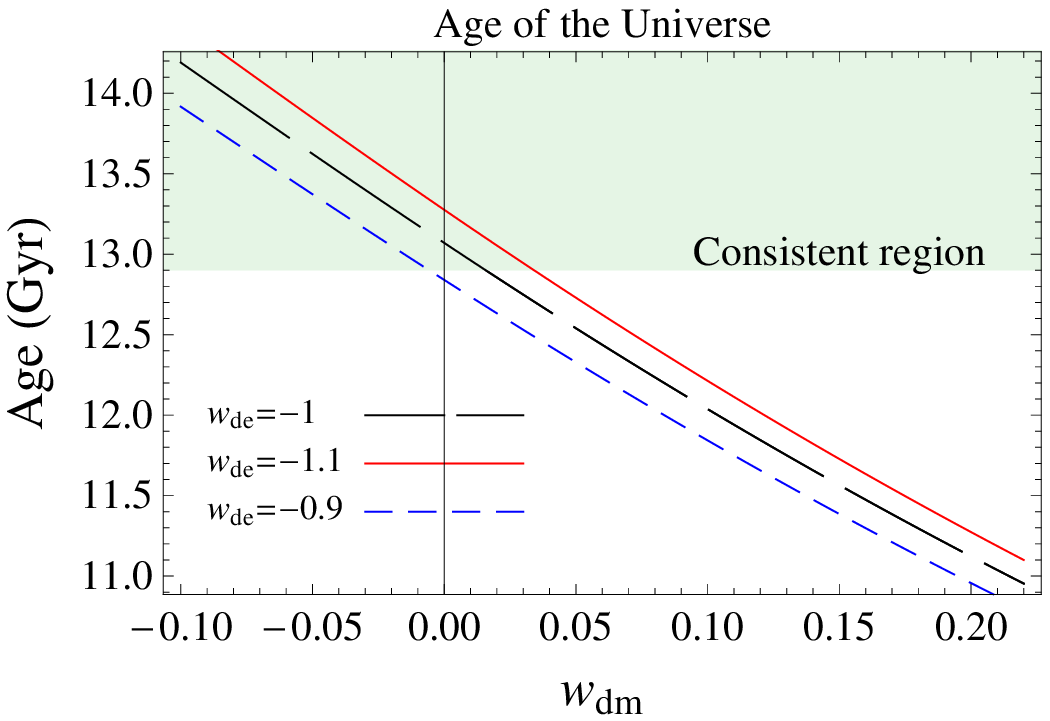}%
\hfill%
\includegraphics[width=7cm]{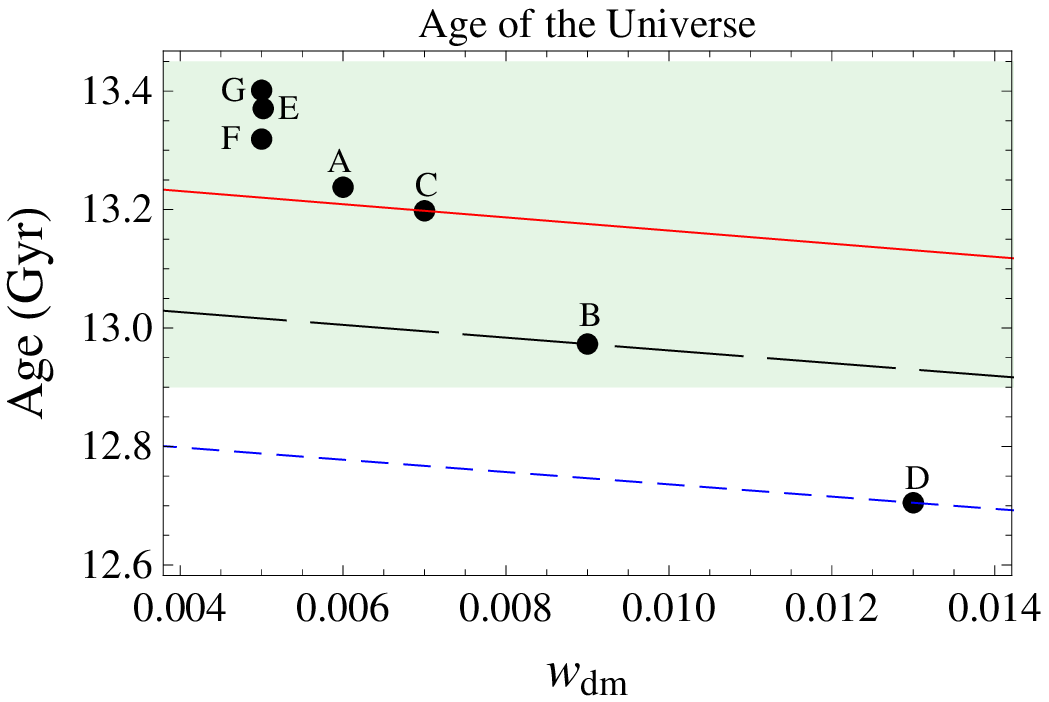}%
\hspace*{\fill}
\caption{Age of the Universe given in gigayears (Gyr) as a function of
$w_{\rm dm}$. The black long dashed, the solid red and the blue short dashed
lines correspond to assume the values of $w_{\rm de} = -1, -1.1, -0.9$
respectively. The right panel corresponds to a zoom in of the left one, where 
the points locate the inferred value of the age of the
Universe when the eq. (\ref{ODEscalefactor}) is evaluated at the best
estimated values for $w_{\rm dm}$ (see section \ref{SectionAgeUniverse}
and table \ref{TableAgeUniverse}). 
The letters that label the points correspond to the values of $(w_{\rm dm},
\mbox{Age}, w_{\rm de},\Omega_{\rm dm0})$ where: A $= (0.006, 13.23, -1.115,
0.23)$, B $= (0.009, 12.97, -1, 0.23)$,  C $= (0.007, 13.19, -1.1, 0.23)$, D
$= (0.013, 12.7, -0.9,  0.23)$, E $= (0.005, 13.37, -1, 0.204)$,
F $= (0.005, 13.31, -1.1, 0.223)$ and G $= (0.005, 13.4, -0.9, 0.184)$.
The shaded area corresponds to the consistent region for the age of the
Universe estimated from the oldest globular clusters (Age$=12.9 \pm 2.9$ Gyr
\cite{AgeUniverse-Carreta2000}).}
\label{PlotAgeUniverse2}
\end{center}
\end{figure}



\begin{table}
  \centering
\begin{tabular}{c | c c c |  c c }

\multicolumn{6}{c}{Age of the Universe}\\

Age (Gyr) & $w_{\rm dm}$ &  $w_{\rm de}$ & $\Omega_{\rm dm0}$ & Table & Point
\\
\hline
\hline

$13.23 \pm 0.06$ & 0.006 &  $-1.115$ &  0.23  & \ref{TableWdmWde} &A \\
$12.97 \pm 0.012$ & 0.009 & $-1$ & 0.23  & \ref{TableWdmAlone} &B \\
$13.19 \pm 0.015$ & 0.007 & $-1.1$ & 0.23  & \ref{TableWdmAlone} &C\\
$12.7 \pm 0.01$ & 0.013 & $-0.9$ & 0.23  & \ref{TableWdmAlone} & D \\
$13.37\pm 0.14$ & 0.005 & $-1$ & $0.204$  & \ref{TableWdmOdm} &E\\
$13.31 \pm 0.12$ & 0.005 & $-1.1$ & $0.223$  & \ref{TableWdmOdm} &F \\
$13.4 \pm 0.12$ & 0.005 & $-0.9$ & $0.184$  & \ref{TableWdmOdm} &G\\
\hline

\end{tabular}
\caption{Age of the Universe given in gigayears (first column) when it is
assumed certain values for $(w_{\rm dm}, w_{\rm de}, \Omega_{\rm dm0})$
shown in the 2nd to 4th columns and that comes from the best estimates of
$w_{\rm dm}$ shown in tables \ref{TableWdmWde}--\ref{TableWdmOdm} (fifth
column). The last column indicates the letters used in figure
\ref{PlotAgeUniverse2} to label those points. $H_0$ is assumed to be 73.8
km/s$\cdot$Mpc. }   
\label{TableAgeUniverse}
\end{table}


	  \subsubsection{The age of the Universe}\label{SectionAgeUniverse}

Using the fact that $H = \dot{a}/a$, we can rewrite the eq.
(\ref{HubbleParameterScaleFactor}) as an ordinary differential equation
(ODE) for the scale factor $a$ in terms of the cosmic time as
\begin{equation}\label{ODEscalefactor}
\frac{da}{dt} - \beta H_0 a \sqrt{\frac{\Omega_{\rm r0}}{a^4} +
\frac{\Omega_{\rm
b0}}{a^3} + \frac{\Omega_{\rm de0}}{a^{3(1+w_{\rm
de})}} + \frac{\Omega_{\rm dm0}}{a^{3(1+w_{\rm
dm})}}} = 0,
\end{equation}

\noindent where $\beta = 1.022729 \times 10^{-3}$ is
introduced to give the units of time in \textit{gigayears} (Gyr) when
the value of the Hubble constant is given in units of km/(s$\cdot$Mpc).
For the conversion of units, we use the values of 1 year = 31558149.8
seconds (a sidereal year)\cite{ParticleDataGroupReview2011} and 1 Mpc $=
3.0856776 \times 10^{19}$ km \cite{ParticleDataGroupReview2011}, so $\beta
= (31558149.8 \times 10^9)/3.0856776$.

We solve numerically the ODE (\ref{ODEscalefactor}) with the initial
condition $a(t=0) = 0$ \footnote{Actually, we used instead $a(t=0) =1
\times 10^{-8}$, to avoid singularities and collapse of the numerical
computing if we set $a=0$ at the eq. (\ref{ODEscalefactor})} and compute
the value $t_{\rm today}$ of the age of the Universe through the condition
$a(t_{\rm today}) = 1$.

Evaluating the numerical solution of the ODE (\ref{ODEscalefactor}) at the
best estimates and assuming the values of $H_0 = 73.8 \pm 2.4$
\cite{Riess:2011yx}, $\Omega_{\rm r0} = 0.0000758 $, $\Omega_{\rm b0} =
0.0458 \pm 0.0016$ \cite{WMAP7yKomatsu2011} we find an age of the Universe.
See table \ref{TableAgeUniverse} and figure \ref{PlotAgeUniverse2}.
From the oldest globular clusters the age of the Universe is constrained to
$12.9 \pm 2.9$ Gyr \cite{AgeUniverse-Carreta2000}.

	    \section{Discussion and Conclusions}

We explored the constraints on the value of the parameter $w_{\rm dm}$ of the
barotropic EoS of the dark matter to investigate the ``warmness'' of the dark
matter fluid.
The model is composed by the dark matter and dark energy fluids in addition to
the radiation and baryon components.
We constrained the value of $w_{\rm dm}$ using the SNe Ia ``Union 2.1'' of
the SCP data set, the three observational ($\mathcal{R}, l_A, z_*$) data from
the CMB given by WMAP-7y, the distance ratio $d_z$ at $z= 0.275$ of BAO and
the Hubble parameter data at different redshifts. 

We calculated the best estimated values for the pair of parameters $(w_{\rm
dm}, w_{\rm de})$,  $(w_{\rm dm}, \Omega_{\rm dm0})$ and also $w_{\rm dm}$
alone, where $w_{\rm de}$ and $\Omega_{\rm dm0}$ are the parameter of the
barotropic EoS of dark energy and the present-day value of the density
parameter of dark matter respectively.

When $w_{\rm dm}$ is estimated together with  $w_{\rm de}$  we found
that the cosmological data prefer the value of $w_{\rm dm} = 0.006 \pm
0.001$, suggesting a \textit{warm} dark matter, and $w_{\rm de}= -1.11 \pm
0.03$ that corresponds to a phantom dark energy, instead a cold
dark matter and a cosmological constant ($w_{\rm dm}=0, w_{\rm de} = -1$).
See table \ref{TableWdmWde} and figures \ref{PlotGroupWdmWde} and
\ref{PlotGroupWdmWdeZoom}.

In order to study the dependence of the estimated value for $w_{\rm dm}$ with
respect to the value of $w_{\rm de}$ of the dark energy, we computed the best
estimate of $w_{\rm dm}$ as the only free parameter but assuming three
different values of $w_{\rm de} =  -1, -1.1, -0.9$.
We found the values of $w_{\rm dm} = 0.009 \pm 0.002$, $0.006
\pm 0.002 $, $0.012 \pm 0.002$ when it is assumed the values of $w_{\rm de} =
-1, -1.1, -0.9$ respectively, where the errors were computed at 3$\sigma$
(99.73\%), so, we found  that $w_{\rm dm} > 0$ with at least 99.73\% of
confidence level. 
Additionally, from these three cases, the assumption of $w_{\rm de} = -1.1$
is the case that allows to fit better the model to data compared with the
other two cases (see table \ref{TableWdmAlone} and figure
\ref{PlotGaussWdmAllTogetherJoinU21FDM}).

When $w_{\rm dm}$ is constrained together with  $\Omega_{\rm dm0}$ we found
that the best fit to data is for $(w_{\rm dm}=0.005 \pm 0.001$, $\Omega_{\rm
dm0} =  0.223 \pm 0.008)$ and with the assumption of $w_{\rm de} = -1.1$,
instead of a cosmological constant (i.e., $w_{\rm de} = -1$).
We found also interesting to notice that the best estimated value of $w_{\rm
dm}$ using all the combined data sets give the same value of $w_{\rm dm}
= 0.005$ independent of the assumed value for $w_{\rm de}$, where the three
cases were $w_{\rm de} =  -1, -1.1, -0.9$ (see the
three rows at the bottom of table \ref{TableWdmOdm}).

In all cases the best fit to data, measured through
the $\chi^2_{\rm d.o.f.}$ magnitude, of the cosmological observations
separately or all together (the joint SNe + CMB + BAO + $H(z)$ data)  
correspond to the case when it is assumed $w_{\rm de} = -1.1$ (phantom dark
energy) instead of a cosmological constant ($w_{\rm de} = -1$) or $w_{\rm de}
= -0.9$.

For the age of the Universe, we found a consistent value for the age when it
is evaluated at the best estimated values for $w_{\rm dm}$, except for the
case when it is assumed $w_{\rm de} = -0.9$. See table
\ref{TableAgeUniverse} and figure \ref{PlotAgeUniverse2}.

On the other hand, Muller \cite{Muller} and more recently Calabrese et al.
\cite{Calabrese2009} investigated the constraints on $w_{\rm dm}$ at
perturbative level comparing with the large scale structure data and CMB
anisotropies. They found the constraints $-0.008 < w_{\rm dm} < 0.0018$ and
$-0.0133 < w_{\rm dm} < 0.0082$ respectively. We find that our results are
comparable and consistent with these ones.

In summary, we found an evidence of a non-vanishing value $w_{\rm dm}$.
From the cosmological observations we found constraints on the values of
$w_{\rm dm}$ around $0.005<w_{\rm dm}<0.01$ suggesting a
\textit{warm} dark matter, independent of assumed value for $w_{\rm de}$, but
where a value $w_{\rm de} < -1$ is preferred by the observations instead of
the $\Lambda$CDM model.
Our constraints on $w_{\rm dm}$ are consistent with perturbative analysis done
in previous works.


\begin{acknowledgments}
N. C. acknowledges the hospitality of the Instituto de F\'{\i}sica
y Matem\'{a}ticas, Universidad Michoacana de San Nicol\'{a}s de
Hidalgo, Morelia, Michoac\'{a}n, M\'{e}xico, where part of this
work was done. A. A. acknowledges the very kind and friendly
hospitality of Prof. Norman Cruz and the Departamento de F\'{\i}sica of the
Universidad de Santiago de Chile where a substantial part of the work was
done. N. C. and A. A. acknowledge the support to this
research by CONICYT through grants N$^{\rm o}$. 1110840 (NC). A. A.
acknowledge the support by SNI-CONACYT and IAC.
U. N. acknowledges the financial support of the SNI-CONACYT, PROMEP-SEP and
CIC-UMSNH. 
\end{acknowledgments}


%

\end{document}